\documentclass{emulateapj}
\usepackage{amsmath}
\usepackage{graphicx}
\usepackage{natbib}
\usepackage[scaled]{helvet}
\usepackage{epsfig}
\usepackage{url}
\bibpunct{(}{)}{;}{a}{}{,}
\newcommand{\kms}{\,km\,s$^{-1}$}
\usepackage{textcomp}
\usepackage{mathpazo}

\usepackage{color}
\usepackage[normalem]{ulem}

\newcommand{\feh}{\ensuremath{\left[{\rm Fe}/{\rm H}\right]}}
\newcommand{\teff}{\ensuremath{T_{\rm eff}}}

\newcommand{\msun}{\ensuremath{\,M_\Sun}}
\newcommand{\rsun}{\ensuremath{\,R_\Sun}}
\newcommand{\lsun}{\ensuremath{\,L_\Sun}}
\newcommand{\mj}{\ensuremath{\,M_{\rm J}}\space}

\newcommand{\ms}{\,m\,s$^{-1}$}
\newcommand{\minus}{\scalebox{0.75}[1.0]{$-$}}

\begin{document}

\title{The Mysterious Dimmings of the T Tauri Star V1334 Tau}
\author{
Joseph E. Rodriguez$^1$, 
George Zhou$^{1}$, 
Phillip A. Cargile$^{1}$, 
Daniel J. Stevens$^{2}$, 
Hugh P. Osborn$^{3}$, 
Benjamin J. Shappee$^{4,5}$, 
Phillip A. Reed$^{6}$, 
Michael B. Lund$^7$, 
Howard M. Relles $^{1}$, 
David W. Latham$^{1}$, 
Jason Eastman $^{1}$, 
Keivan G. Stassun$^{7,8}$, 
Allyson Bieryla$^{1}$, 
Gilbert A. Esquerdo$^{1}$, 
Perry Berlind$^{1}$, 
Michael L.\ Calkins$^{1}$, 
Andrew Vanderburg$^{1}$, 
Eric Gaidos$^{9,10}$, 
Megan Ansdell$^{11}$, 
Robert J. Siverd$^{12}$,
Thomas G. Beatty$^{13,14}$, 
Christopher S. Kochanek$^{2,15}$, 
Joshua Pepper$^{16}$, 
B. Scott Gaudi$^{2}$, 
Richard G. West$^{3}$, 
Don Pollacco$^{3}$, 
David James$^{17}$, 
Rudolf B. Kuhn$^{18}$,
Krzysztof Z. Stanek$^{2,15}$, 
Thomas W.-S. Holoien$^{2,15,19}$, 
Jose L. Prieto$^{20,21}$, 
Samson A. Johnson$^{2}$, 
Anthony Sergi$^{22}$, 
Nate McCrady$^{22}$, 
John A. Johnson$^{1}$, 
Jason T. Wright$^{13,14}$, 
Robert A. Wittenmyer$^{23,24}$, 
Jonathan Horner$^{23,24}$
}

\affil{$^{1}$Harvard-Smithsonian Center for Astrophysics, 60 Garden St, Cambridge, MA 02138, USA}
\affil{$^{2}$Department of Astronomy, The Ohio State University, Columbus, OH 43210, USA}
\affil{$^{3}$Department of Physics, University of Warwick, Gibbet Hill Road, Coventry, CV4 7AL, UK}
\affil{$^{4}$Carnegie Observatories, 813 Santa Barbara Street, Pasadena, CA 91101, USA}
\affil{$^{5}$Hubble, Carnegie-Princeton Fellow}
\affil{$^{6}$Department of Physical Sciences, Kutztown University, Kutztown, PA 19530, USA} 
\affil{$^{7}$Department of Physics and Astronomy, Vanderbilt University, 6301 Stevenson Center, Nashville, TN 37235, USA}
\affil{$^{8}$Department of Physics, Fisk University, 1000 17th Avenue North, Nashville, TN 37208, USA}
\affil{$^{9}$Department of Geology and Geophysics, University of Hawai`i at Mānoa, Honolulu, HI 96822}
\affil{$^{10}$Center for Space and Habitability, University of Bern, CH-3201 Bern, Switzerland}
\affil{$^{11}$Institute for Astronomy, University of Hawai`i at Manoa, Honolulu, HI 96822, USA}
\affil{$^{12}$Las Cumbres Observatory Global Telescope Network, 6740 Cortona Dr., Suite 102, Santa Barbara, CA 93117, USA}
\affil{$^{13}$Department of Astronomy \& Astrophysics, The Pennsylvania State University, 525 Davey Lab, University Park, PA 16802}
\affil{$^{14}$Center for Exoplanets and Habitable Worlds, The Pennsylvania State University, 525 Davey Lab, University Park, PA 16802}
\affil{$^{15}$Center for Cosmology and AstroParticle Physics (CCAPP), The Ohio State University, 191 W.\ Woodruff Ave., Columbus, OH 43210, USA}
\affil{$^{16}$Department of Physics, Lehigh University, 16 Memorial Drive East, Bethlehem, PA 18015, USA}
\affil{$^{17}$Astronomy Department, University of Washington, Box 351580, Seattle, WA 98195, USA}
\affil{$^{18}$South African Astronomical Observatory, PO Box 9, Observatory 7935, South Africa}
\affil{$^{19}$US Department of Energy Computational Science Graduate Fellow}
\affil{$^{20}$N´ucleo de Astronom´ıa de la Facultad de Ingenier´ıa, Universidad Diego Portales, Av. Ej´ercito 441, Santiago, Chile}
\affil{$^{21}$Millennium Institute of Astrophysics, Santiago, Chile}
\affil{$^{22}$Department of Physics and Astronomy, University of Montana, Missoula, MT 59812, USA}
\affil{$^{23}$Computational Engineering and Science Research Centre, University of Southern Queensland, Toowoomba, QLD 4350, Australia}
\affil{$^{24}$School of Physics and Australian Centre for Astrobiology, University of New South Wales, Sydney 2052, Australia}
\shorttitle{V1334 Tau}

\begin{abstract}
We present the discovery of two extended $\sim$0.12 mag dimming events of the weak-lined T-Tauri star V1334. The start of the first event was missed but came to an end in late 2003, and the second began in February 2009, and continues as of November 2016. Since the egress of the current event has not yet been observed, it suggests a period of $>$13 years if this event is periodic. Spectroscopic observations suggest the presence of a small inner disk, although the spectral energy distribution shows no infrared excess. We explore the possibility that the dimming events are caused by an orbiting body (e.g. a disk warp or dust trap), enhanced disk winds, hydrodynamical fluctuations of the inner disk, or a significant increase in the magnetic field flux at the surface of the star. We also find a $\sim$0.32 day periodic photometric signal that persists throughout the 2009 dimming which appears to not be due to ellipsoidal variations from a close stellar companion. High precision photometric observations of V1334 Tau during K2 campaign 13, combined with simultaneous photometric and spectroscopic observations from the ground, will provide crucial information about the photometric variability and its origin.

\end{abstract}

\shortauthors{Rodriguez et al.}

\section{Introduction}

The circumstellar environments of young stellar objects (YSOs) are complex because a large number of processes drive their evolution. The observed properties of YSOs involve accretion onto the star, dispersion by stellar winds and radiation, magnetic fields, outflows, and in many cases stellar companions. Many of these characteristics can manifest themselves through disk substructures, gradients, and other disk properties that could reveal the mechanisms that influence planet formation. Although high spatial resolution imaging and interferometry can now resolve the circumstellar environment of some young stars, we are not yet able to probe the innermost region for most stars ($<$10 AU) where a large fraction of the exoplanets reside.

First classified by \citet{Herbst:1994}, YSOs are known to vary in the optical as a result of material accreting onto the stellar surface and/or circumstellar extinction. This variability can vary in duration, depth, and periodicity. Some YSOs showing non-periodic, large ($>$1 mag) and long (months to years) dimmings and are referred to as UX Orionis stars (UXors). Proposed causes are over-dense regions in the nearly edge-on circumstellar disk surrounding the host star \citep{Wenzel:1969, Grinin:1988, Voshchinnikov:1989, Grinin:1998, Grady:2000}, hydrodynamical fluctuations of the inner disk \citep{Dullemond:2003}, or strong disk winds \citep{Petrov:2007}. 

Shorter duration (days to weeks) photometric variability of YSOs has been seen in both the optical and infrared but the cause is unclear. This variability could be caused by accretion of circumstellar material onto the stellar surface causing hot spots, circumstellar extinction, or orbiting dust clumps \citep{Herbst:1994, Stassun:1999, Bouvier:2007}. These stars are typically referred to as ``dippers'', and although the variability is shorter in duration, the dimming can have a depth of up to $\sim$1 mag. Dippers have a lower amplitude in the infrared than the optical, supporting the interpretation that they are caused by circumstellar dust obscuration \citep{Cody:2014}.

A unique way to study protoplanetary formation and evolution is to observe YSOs that exhibit significant large dimming events ($>$10\% depth and months to years in duration) caused by a portion of the star's circumstellar disk. While these events are quite rare, they provide a powerful tool for studying the circumstellar environment of YSOs. Using high cadence photometric observations from the Kilodegree Extremely Little Telescope (KELT, \citealp{Pepper:2007,Pepper:2012}), we have been conducting the Disk Eclipse Search with KELT (DESK) survey to identify and characterize these rare systems \citep{Rodriguez:DESK}. The DESK survey has discovered and analyzed previously unknown disk eclipsing events around the stars RW Aurigae \citep{Rodriguez:2013, Rodriguez:2016A}, V409 Tau and AA Tau \citep{Rodriguez:2015}, TYC 2505-672-1 \citep{Rodriguez:2016B}, and DM Ori \citep{Rodriguez:2016C}. 

\begin{figure*}[!ht]
\vspace{0.3in}
\includegraphics[width=0.9\linewidth, trim = 0 6.3in 0 0]{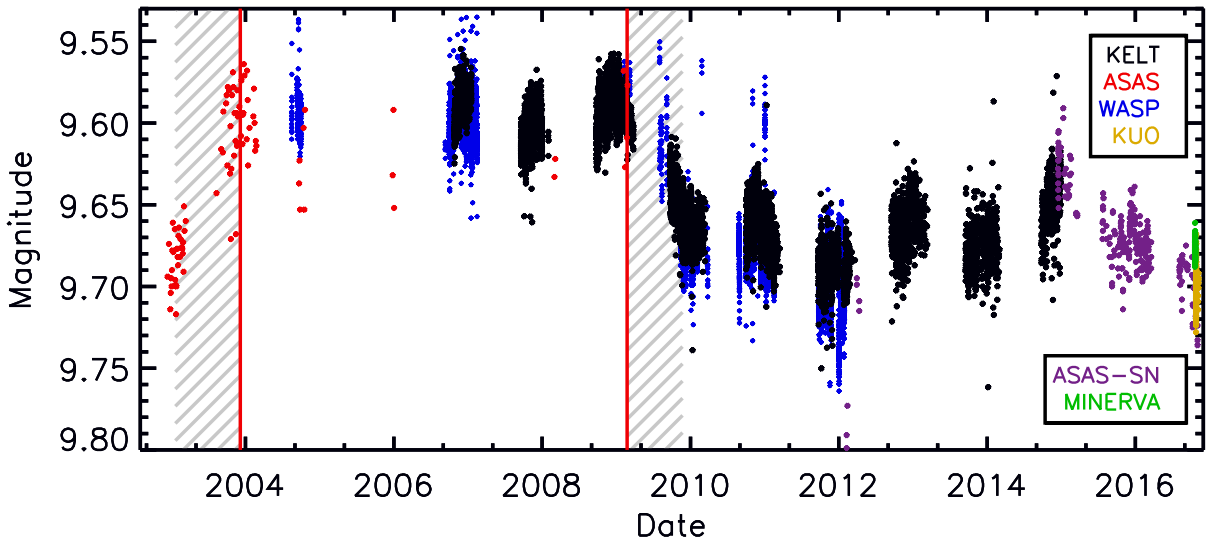}
\caption{The KELT (black), ASAS (red), WASP (blue), ASAS-SN (purple), KUO (yellow), and MINERVA (green) photometric observations of V1334 Tau from 2002 until late 2016. Only the $V$-band observations from KUO and MINERVA are shown. The red vertical lines represent the estimated end of the pre-2004 events and the start of the 2009 event. The grey shaded region represents the estimated egress/ingress for each event described in \S\ref{sec:results}. The intercalibration of the data sources is described in \S\ref{sec:results}.}
\label{figure:FullLC}
\end{figure*}

Interestingly, not all of these discoveries are YSOs. TYC 2505-672-1 is an evolved M giant being eclipsed every $\sim$69 years by a companion with a disk. This system is very similar to the archetype, $\epsilon$ Aurigae, an F-giant being eclipsed every $\sim$27 years by a small star embedded in a circumstellar disk \citep{Carroll:1991}. Interferometric observations by the Georgia State University’s Center for High Angular Resolution Astronomy (CHARA) during the last eclipse of $\epsilon$ Aur confirmed this enterpretation by imaging the companion and disk crossing the star \citep{Kloppenborg:2010}. These evolved systems still have remnant circumstellar material in their systems. Since these discoveries are all at different ages and stages of disk evolution, each system provides information about different stages of protoplanetary formation and evolution which may in turn provide insight into the large diversity of exoplanetary systems.  

In this paper, we present new photometric and spectroscopic observations for the Weak-lined T Tauri Star (WTTS) V1334 Tau. Here we find two dimming events of V1334 Tau, one that ended in late 2003 and another that began in 2009 and extends to the present date. The known characteristics of the V1334 Tau system are described in \S2. The photometric observations are presented in \S3. We present our estimate of the stellar parameters of V1334 Tau A in \S4. We describe the spectroscopic analysis and the long-term photometric variability in \S5, discuss possible dimming mechanisms in \S6, and summarize our results in \S7.



\section{The V1334 Tau System}
V1334 Tau was previously identified as a Weak-lined T Tauri Star (WTTS) with a K1 spectral type by \citet{Wichmann:1996} using observations from the ROSAT All-Sky-Survey \citep{Zimmermann:1993} combined with optical spectra. More recently, V1334 Tau was re-classified as a G2 star \citep{Nguyen:2012}. \citet{Wichmann:2000} determined that V1334 Tau has a radius of 2.73 $\rsun$, a mass of 2.05 $\msun$, and an age of 3.83$\times$10$^{6}$ yr. V1334 Tau has a proper motion of $\mu_{\alpha}$=8.7$\pm$ 0.8 and $\mu_{\delta}$= $\minus$24.7 $\pm$ 0.7	mas yr$^{-1}$ which is in agreement with the known motion of the Taurus-Aurigae association ($\mu_{\alpha}$=7.2 and $\mu_{\delta}$=-20.9, \citealp{Bertout:2006}). Using speckle observations, \citet{Kohler:1998} identified a close companion to V1334 Tau (1\farcs631, $\Delta$K = 3.42). A near-IR high-spatial-resolution survey for low-mass companions in Taurus by \citep{Daemgen:2015} found two additional nearby companion candidates at 0\farcs106 with a $\Delta$K = 1.68$\pm$0.05 and 6\farcs969 with a $\Delta$K = 12.5$\pm$0.3.

\begin{table*}
\centering
\caption{Stellar Properties and photometric measurments of V1334 Tau obtained from the literature.}
\label{tbl:Host_Lit_Props}
\begin{tabular}{llccc}
   \hline
  \hline
\hline
  Parameter & Description & Value & Source & Reference(s) \\
Names& 					&  V1334 Tau	& 		&			\\
			& 					& HD 290380		& 		&			\\
			&					&TYC 1839-643-1&		&			\\
			&					&2MASS J04445445+2717454		&		&			\\
			&					&				&		&			\\
$\alpha_{J2000}$	&Right Ascension (RA)& 04:44:54.454			& Tycho-2	& \citet{Hog:2000}	\\
$\delta_{J2000}$	&Declination (Dec)& +27:17:45.23			& Tycho-2	& \citet{Hog:2000}	\\
B$_T$			&Tycho B$_T$ magnitude& 10.909 $\pm$ 0.061		& Tycho-2	& \citet{Hog:2000}	\\
V$_T$			&Tycho V$_T$ magnitude& 9.895 $\pm$ 0.038		& Tycho-2	& \citet{Hog:2000}	\\
			&					&				&		&			\\
FUV         & Far UV magnitudes & 20.05 & GALEX & \citet{Gomez:2015} \\
NUV         & Near UV magnitudes & 15.63 & GALEX & \citet{Gomez:2015} \\
            &                   &               &       &           \\
$V$ & Johnson V&9.612 $\pm$ 0.04& APASS 	& \citet{Henden:2016}	\\
$B$ &Johnson B&10.511  $\pm$  0.063& APASS 	& \citet{Henden:2016}	\\
$g'$&   Sloan g'	&10.312  $\pm$  0.045& APASS 	& \citet{Henden:2016}	\\
$r'$ &  Sloan r'	&9.361  $\pm$  0.06& APASS 	& \citet{Henden:2016}	\\
$i'$ &  Sloan i'	&8.923 $\pm$  0.05& APASS 	& \citet{Henden:2016}	\\
			&					&				&		&			\\
J			&2MASS magnitude& 7.734  $\pm$ 0.02		& 2MASS 	& \citet{Cutri:2003}	\\
H			&2MASS magnitude& 7.281 $\pm$ 0.04	& 2MASS 	& \citet{Cutri:2003}	\\
K$_{s}$	    &2MASS magnitude& 7.154  $\pm$ 0.02		& 2MASS 	& \citet{Cutri:2003}	\\
			&					&				&           &			\\
\textit{WISE1}		&WISE passband& 7.017  $\pm$ 0.033		& WISE 		&\citet{Cutri:2012}	\\
\textit{WISE2}		&WISE passband&  7.060  $\pm$ 0.018		& WISE 		& \citet{Cutri:2012}\\
\textit{WISE3}		&WISE passband& 7.048 $\pm$ 0.018		& WISE 		& \citet{Cutri:2012}	\\
\textit{WISE4}		&WISE passband& 7.058 $\pm$ 0.108       & WISE 		& \citet{Cutri:2012}	\\
			&					&				&		&			\\
$\mu_{\alpha}$		& Proper Motion in RA (mas yr$^{-1}$)	& 8.7$\pm$ 0.8	& NOMAD		& \citet{Zacharias:2004} \\
$\mu_{\delta}$		& Proper Motion in DEC (mas yr$^{-1}$)	&  $\minus$24.7 $\pm$ 0.7	& NOMAD		& \citet{Zacharias:2004} \\
			&					&				&		&			\\
$RV$\dotfill & Systemic radial \hspace{9pt}\dotfill  & $\sim$19  & & this work \\
     & \hspace{3pt} velocity (\kms)  & & \\
$U^{*}$\dotfill & Space motion (\kms)\dotfill & $-12.1\pm 0.6 $  & & this work \\
$V$\dotfill       & Space motion (\kms)\dotfill & $24.3\pm 1.6$ &  & this work \\
$W$\dotfill       & Space motion (\kms)\dotfill & $16.2\pm 2.4$ &  & this work \\
			
\hline
\hline
\end{tabular}

\begin{flushleft}
 \footnotesize \textbf{\textsc{NOTES}}\\
 \footnotesize The photometry shown is used in \S \ref{sec:SED} to determine the stellar parameters.
\footnotesize $U^{*}$ is positive in the direction of the Galactic Center. UVW analysis suggests a 98.4\% chance of V1334 Tau being in the thin disk according to the classification system of \citet{Bensby:2003}. The UVW kinematics used the peculiar velocity of the Sun with respect local standard rest from \citet{Coskunoglu:2011}. The UVW analysis and distance from \S \ref{sec:SED} are consistent with this being a member of the Taurus-Aurigae association.
\end{flushleft}
\end{table*}

\section{Observations}
Over the past decade multiple ground-based photometric surveys have observed V1334 Tau. None of the photometric observations resolve the fainter components in the system.  All the follow-up photometry shown in Figure \ref{figure:FullLC} is available in machine-readable form in the online journal.

\subsection{KELT}
With the primary goal of discovering transiting planets around bright host stars ($8<V<11$), the KELT project is a ground-based photometric survey covering $>$70\% of the entire sky on a 10-20 minute cadence \citep{Pepper:2007, Pepper:2012}. The project consists of two telescopes, KELT-North in Sonoita, AZ and KELT-South in Sutherland, South Africa. Each telescope has a Mamiya 645-series wide-angle lens with a 42mm aperture and a 80mm focal length (f/1.9), and observes in a broad R-band filter. The telescopes have a $26^{\circ}$ $\times$ $26^{\circ}$ field-of-view (FOV), and a 23$\arcsec$ pixel scale. V1334 Tau is located in KELT-North field 03 (field center at $\alpha$ = 3h 58m 12s, $\delta$ = $31^{\circ}$ 39$\arcmin$ 56.16$\arcsec$). The KELT-North telescope observed V1334 Tau from UT 2006 October 26 to UT 2013 March 13, obtaining 9186 observations after processing. For a detailed description of the KELT data acquisition and reduction process, see \citet{Siverd:2012}. The median per-point error is 0.005 mag. 

\subsection{All-Sky Automated Survey for SuperNovae (ASAS-SN)}
Designed to discover nearby supernovae, the All-Sky Automated Survey for SuperNovae (ASAS-SN) is photometrically monitoring the observable sky every two days down to $V$ $\sim$ 17 \citep{Shappee:2014}. The survey has two sites, Cerro Tololo InterAmerican Observatory (CTIO) in Chile and Mount Haleakala in Hawaii. Each site is hosted by the Las Cumbres Observatory Global Telescope Network \citep{Brown:2013} and has two units with four 14 cm Nikon telephoto lenses and a 2k $\times$ 2k thinned CCD, with a $4.5\times4.5$ degree FOV and a 7$\farcs$8 pixel scale. V1334 Tau was observed 532 times from UT 2012 January 20 until UT 2016 October 14. The median per-point error is 0.004 mag.

\subsection{SuperWASP}
The Wide Angle Search for Planets (SuperWASP) is a high-cadence photometric survey designed to detect transiting exoplanets. Observing in a broad filter (centered at 550 nm), the survey has two observing locations, one at the Roque de los Muchachos Observatory on La Palma (WASP-North) and the other located at the South African Astronomical Observatory (WASP-South). With 8 cameras at each location, each using a 200 mm f/1.8 lens and a 2048$\times$2048 pixel CCD, SuperWASP is able to monitor large portions of the sky at a very high cadence (minutes). Each camera has a $7.8^{\circ}$ $\times$ $7.8^{\circ}$ FOV, and a 13$\farcs$7 pixel scale. V1334 Tau was observed in two separate seasons in the publicly released observations: UT 2004 August 15 to UT 2004 September 30 and UT 2006 September 11 until UT 2007 February 15\footnote{http://exoplanetarchive.ipac.caltech.edu/}. All observations from WASP prior to 2006 are unfiltered and after 2006, the observations are in a broad $V$-band filter. A 3$\sigma$ clipping around the median was done to remove large outliers. V1334 Tau was also observed from UT 2009 February 08 until UT 2012 January 29. Since the data reduction of the public SuperWASP data was performed differently than the more recent WASP observations, the two data sets were kept separate throughout the analysis described here. Additionally, we binned the second SuperWASP data set in 1-hour bins and calculated the standard deviation of each bin. All data points within bins which had a standard deviation greater than 2\% were then discarded. After processing and cleaning, 14339 observations remained and were used in this work. The median per-point error is 0.009 mag for the public observations and 0.003 mag for the recent observations.

\subsection{All-Sky Automated Survey (ASAS)}
Designed to discover and catalog variable stars over the entire sky, the All-Sky Automated Survey (ASAS) observes all stars brighter than $V$=14. For a detailed description of the observing strategy and data reduction, see \citet{Pojamanski:1997}. ASAS uses two locations, Las Campanas, Chile and Haleakala, Maui. Each observing location observed in the $V$ and $I$ bands simultaneously using two wide-field Minolta 200/2.8 APOG telephoto lenses with a 2K$\times$2K Apogee CCD. Each telescope has a $8.8^{\circ}$ $\times$ $8.8^{\circ}$ FOV. ASAS observed V1334 Tau in the $V$-band from UT 2002 December 13 until UT 2009 February 25. 93 observations were obtained over this period. The median per-point error is 0.034 mag.

\subsection{MINERVA}
The MINERVA Project has four 0.7m PlaneWave CDK-700 telescopes located on Mt. Hopkins, AZ, at the Fred L. Whipple Observatory, capable of both millimag photometry and precision radial velocity measurements \citep{swift:2015}. On the night of UT 2016 October 21, two telescopes simultaneously observed V1334 Tau using an Andor iKON-L 2048$\times$2048 detector with a 20\farcm9 $\times$ 20\farcm9 FOV and a plate scale of 0$\farcs$6 pixel$^{-1}$. One telescope observed in the V-band with 200 exposures and achieved a per-point RMS scatter of 0.0009 mag. The other observed in the B band with 140 exposures and achieved a per-point RMS scatter of 0.0015 mag.

\subsection{Kutztown University Observatory}
Using the 0.61 m Ritchey-Chr\'{e}tien optical telescope at the Kutztown University Observatory (KUO) in Kutztown, Pennsylvania, V1334 Tau was observed in the $B$ and $V$-bands on UT 2016 October 26, UT 2016 November 05 and 07. The observing setup uses a $3072 \times 2048$ CCD that has a $19\farcm5 \times 13\farcm0$ FOV with a 0\farcs38 pixel$^{-1}$. Standard aperture photometry was used, and the instrumental flux was color-corrected using several Landolt standard fields. In total, KUO obtained 160 $B$-band and 186 $V$-band observations. The median per point a per-point RMS is 0.0007 mag in the $B$-band and 0.0006 $V$-band.

\subsection{Spectroscopic Follow-up}
We obtained spectroscopic observations of V1334 Tau with the Tillinghast Reflector Echelle Spectrograph (TRES) on the 1.5\,m telescope at the Fred Lawrence Whipple Observatory, on Mount Hopkins, Arizona, USA from UT 2016 October -- December. We obtained nineteen 10 minute exposures over the span of the two months, with fourteen observations occurring on UT 2016 December 09 and 10. The spectra were obtained at a resolution of $\lambda / \Delta \lambda \equiv R = 44000$ over the wavelength range of $3900-9100$\AA. The spectra are consistent with a $\sim$5500\,K dwarf and our estimate from \S \ref{sec:SED}, rotationally broadened by $77.6,\mathrm{km\,s}^{-1}$. We estimate the absolute radial velocity of the V1334 Tau system to be $\sim$19 \kms which is in agreement for the known absoluted radial velocity of the Taurus-Auriga association (15 \kms, \citealp{Bertout:2006}). The TRES observations show no large ($>$1 \kms) radial velocity changes.

\begin{figure}[!ht]
\vspace{0.3in}
\centering\includegraphics[width=0.99\linewidth, angle = 0, trim = 0 0 0 1in]{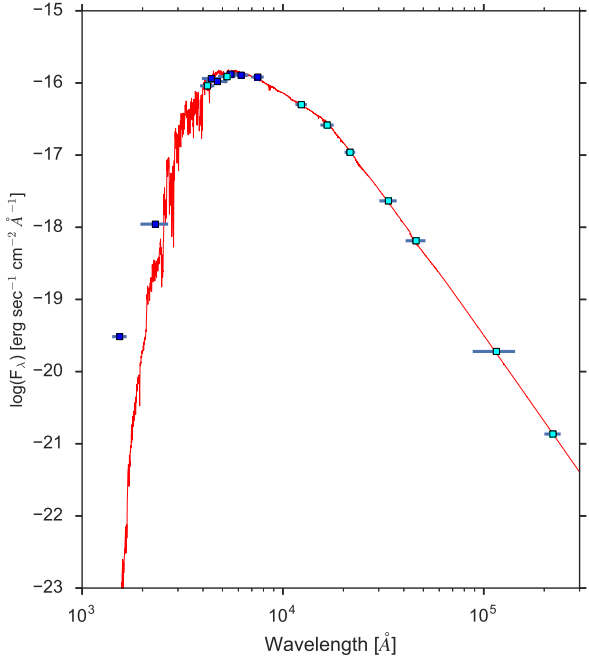}   
\caption{Spectral energy distribution of V1334 Tau along with a model using the most probable stellar parameters from our {\sc MINES}{\lowercase {\it weeper}} analysis. The light blue points show the Tycho-2 B$_{T}$, V$_{T}$, 2MASS J, H, K$_{s}$, and WISE 1--4 photometric observations included in the fit. The dark blue points are the GALEX and APASS photometry that were not used in the fit. The UV excess is likely due to chromospheric activity.}
\label{SED_Fit}
\end{figure}

\section{Determination of Stellar Parameters}
\label{sec:SED}
In order to determine the stellar parameters of V1334 Tau, we use {\sc MINES}{\lowercase {\it weeper}}, a newly developed Bayesian approach to determining stellar parameters. A full description of {\sc MINES}{\lowercase {\it weeper}} is given in Cargile et al. (in prep); here we provide a brief summary of the method. {\sc MINES}{\lowercase {\it weeper}} uses nested importance sampling to determine posterior probability distributions for physical parameters inferred from stellar evolution models. {\sc MINES}{\lowercase {\it weeper}} uses a modified version of the nestle.py code \footnote{http://kbarbary.github.io/nestle/} to perform multi-nested ellipsoid sampling based on the algorithm described in \citet{Feroz:2009}. Multi-nested sampling is uniquely suited to the problem of modeling parameters from stellar evolution models due to its ability to efficiently sample the multi-modal likelihood surfaces, frequently found when modeling stars with stellar isochrones. {\sc MINES}{\lowercase {\it weeper}} uses the most recent release of the MIST stellar evolution models \citep{Choi:2016}, and an optimized interpolation schema based on the recommendations of \citet{Dotter:2008}. {\sc MINES}{\lowercase {\it weeper}} can use both photometric and spectroscopic measurements as well as a wide range of prior probability distributions. The {\sc MINES}{\lowercase {\it weeper}} inference results in posterior probability distributions for the fundamental MIST model parameters: equal evolutionary points (EEP), input stellar metallicity, and stellar age. Using these posterior distributions, {\sc MINES}{\lowercase {\it weeper}} then generates posterior distributions for any other stellar parameters predicted by the MIST models such as the mass, radius, T$_{\rm eff}$, luminosity, and surface abundances.

\begin{table}
 \small
 \caption{Determined Stellar parameters for V1334 Tau A and 68\% confidence interval values}
 \label{tab:parameters}
 \begin{tabular}{ccc}
    \hline
    \hline
Parameter &  Description & Determined Value \\
    \hline
Age & Stellar Age& 1$^{+1}_{-1}$ Myr      \\
M$_{\star}$ & Stellar Mass       & 1.674$^{+0.229}_{-0.279}$ \msun      \\
R$_{\star}$ & Stellar Radius       & 2.619$^{+0.597}_{-0.304}$ \rsun      \\
log(L$_{\star}$) &   Log Stellar Luminosity     & 0.658$^{+0.103}_{-0.214}$  \lsun    \\
\teff &  Effective Temperature      & 5015$^{+93}_{-154}$ K      \\
log(g) &   Surface Gravity     & 3.767$^{+0.114}_{-0.085}$ cgs \\
\feh$_{initial}$ &   Metallicity at formation     & $-$0.116$^{+0.081}_{-0.164}$ \\
\feh$_{surface}$ &   Surface Metallicty      & $-$0.077$^{+0.083}_{-0.171}$   \\
Distance &  \dotfill      & 132$^{+32}_{-14}$ Pc      \\
Av &    Extinction    & 0.438$^{+0.092}_{-0.160}$ mag      \\
     \hline
    \hline
 \end{tabular}
\begin{flushleft}
  \end{flushleft}
\end{table}

We use {\sc MINES}{\lowercase {\it weeper}} to infer the physical properties (e.g., mass, radius, age, etc.) of V1334 Tau, as well as its distance and extinction (A$_{V}$). We modeled all the available broadband optical and IR photometry (see Table \ref{tbl:Host_Lit_Props}) except the GALEX UV and APASS photometry. We are hesitant to include the GALEX data due to the likely presence of chromospheric emission in such a young rapidly rotating star. The APASS data were not included due to our past experience with unaccounted for zero-point offsets seen in this survey's photometry (Cargile et al. in prep). We applied a Gaussian distance prior centered at 140 pc with a FWHM of 40pc \citep{Torres:2007,Torres:2009}, and a uniform age prior (0 $-$ 50 Myr). 

\begin{figure*}[!ht]
\vspace{0.3in}
\centering\includegraphics[width=0.9\linewidth, angle = 0, trim = 0.5in 0 1in 0]{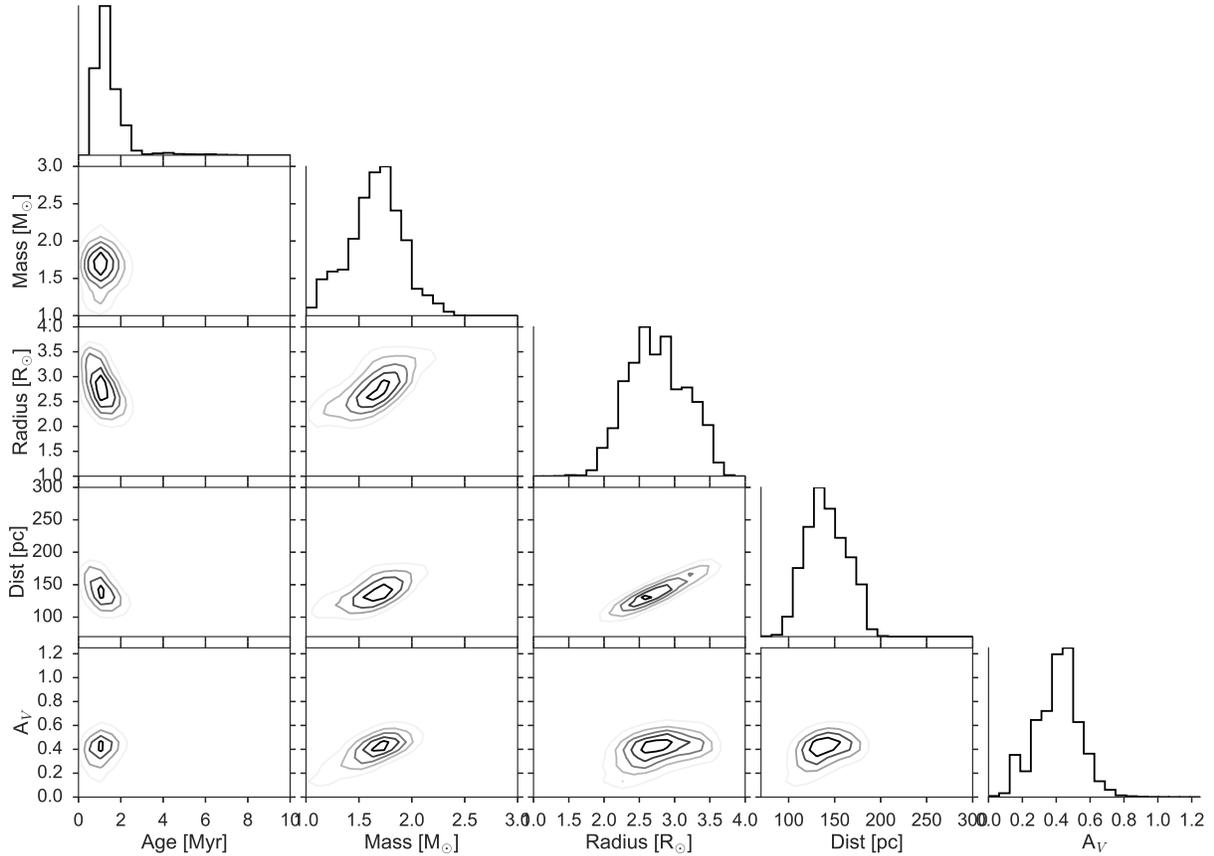}   
\caption{Posterior probability distributions of the stellar parameters for V1334 Tau based on our {\sc MINES}{\lowercase {\it weeper}} analysis. The parameters in Table \ref{tab:parameters} are the maximum posterior probabilities of these posteriors along with their 68\% confidence intervals.}
\label{Triangle}
\end{figure*}

The spectral energy distribution (SED) for V1334 Tau does not have an infrared (IR) excess indicative of a circumstellar disk (See Figure \ref{SED_Fit}), but there is an excess in the ultra-violet (UV) when compared to the MIST stellar SED model. In Figure \ref{Triangle}, we show the posterior probability distributions for the age, mass, radius, distance, and A$_{V}$. Our inferred distance and age agrees with standard estimates for solar-type stars in the Taurus star-forming region, $\sim$130$\pm$25 pc \citep{Torres:2007,Torres:2009} and $\sim$1 Myr \citep{Briceno:2002,Luhman:2003} -- suggesting that V1334 Tau appears to be a {\it bona fide} member of this stellar association. Our models suggest that the mass of V1334 Tau is lower than that proposed by \citet{Wichmann:2000} but within 1$\sigma$ (See Table \ref{tab:parameters} for our determined stellar parameters).


\section{Results}
\label{sec:results}
In this section, we examine the available photometry. We explore different interpretations for the long-duration dimming events observed in \S7. All photometric observations displayed in Figure \ref{figure:FullLC} are in the $V$-band (ASAS, ASAS-SN, KUO, and MINERVA) or a broader filter (KELT and WASP). To place all photometric data on the same scale, we use the KUO observations as the photometric standard and apply a vertical offset to each data set to align them where they overlap. We do not otherwie correct for the filter differences.

\subsection{Out-of-Dimming Variability}
\label{sec:oot_phot-var}
\begin{figure*}[!ht]
\centering
\includegraphics[width=0.99\linewidth]{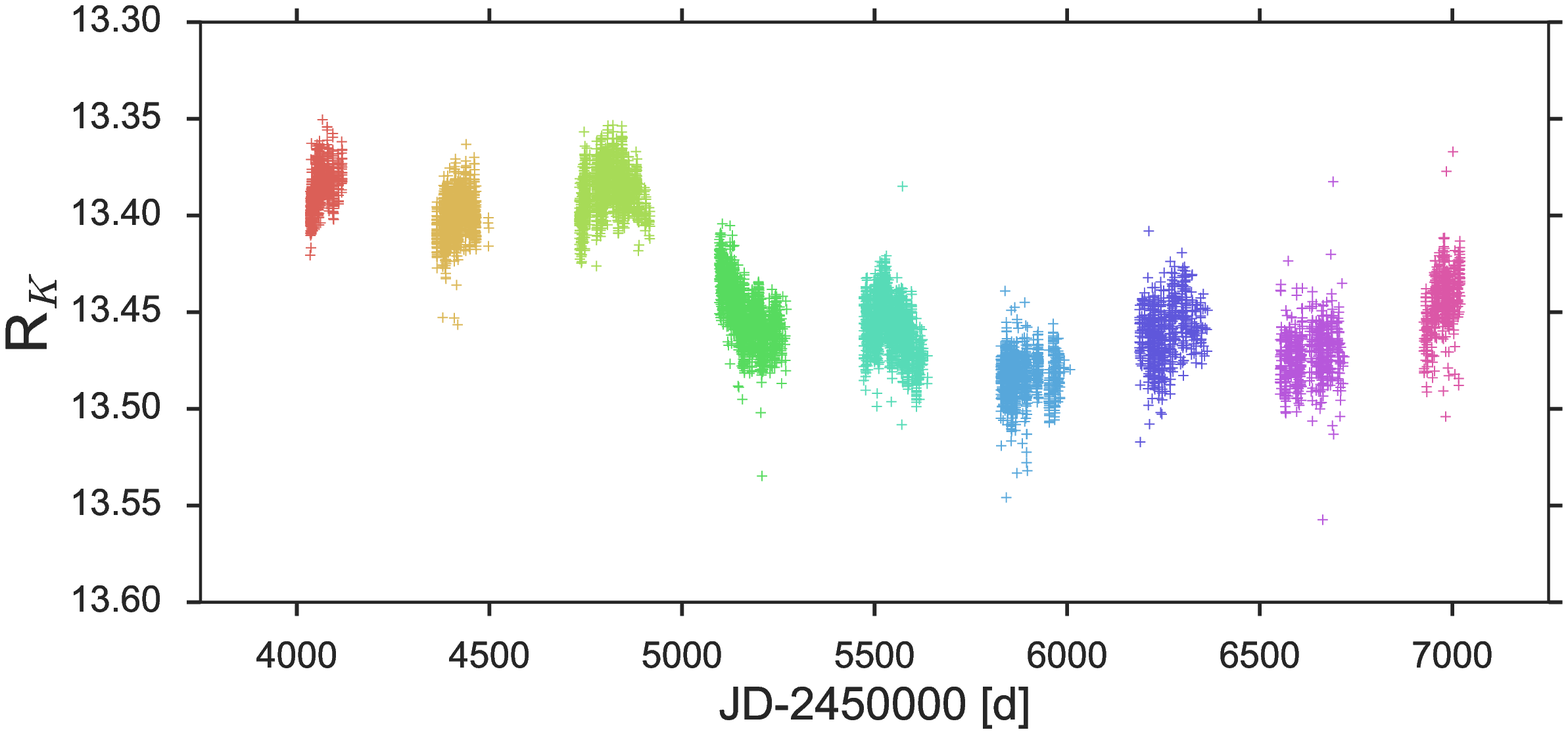}\\
\begin{tabular}{cc}
    \includegraphics[width=0.49\linewidth]{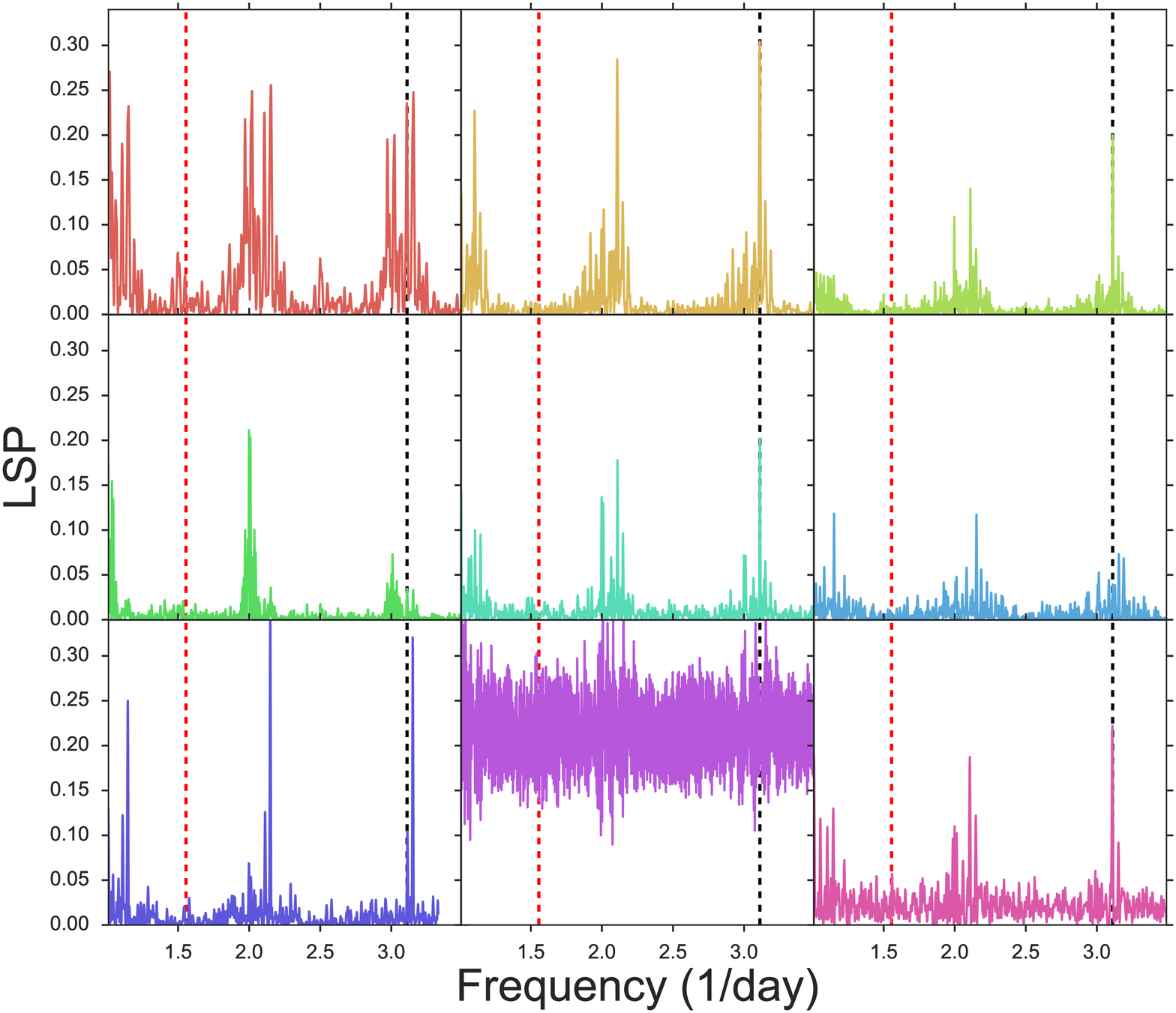}
    \includegraphics[width=0.49\linewidth]{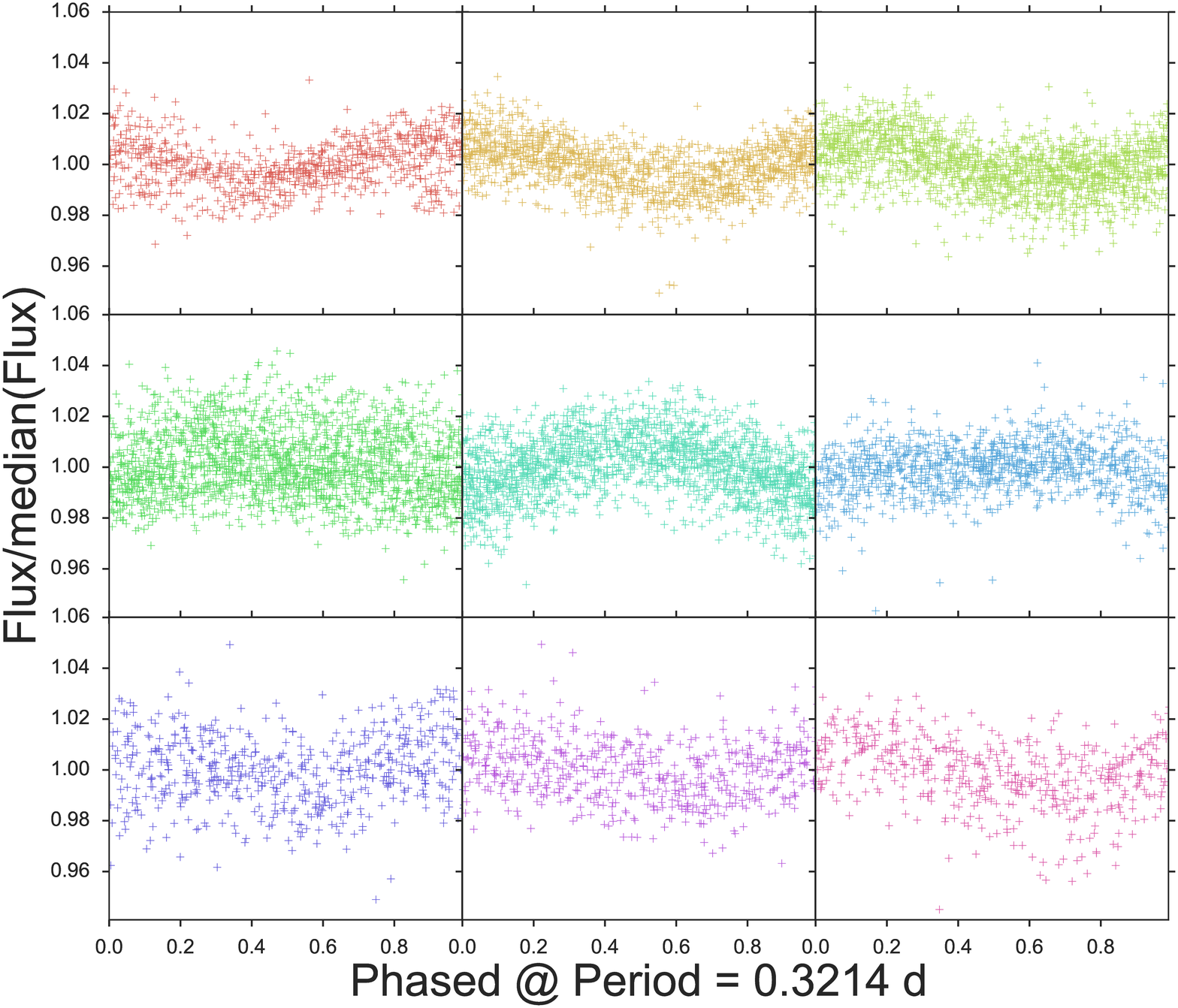}\\  
\end{tabular}
\caption{(Top) The 9 year light curve of V1334 Tau color coded by observing season. (Bottom-left) The LS periodograms for each KELT season. The vertical lines correspond to the 0.32 day period estimate (black) and twice that period (Red). (Bottom-right) Each season of KELT data phase-folded to the 0.32 day period. Note that the phase and amplitude of the 0.32 day signal is changing over the 9 observing seasons from KELT. }
\label{figure:phased}
\end{figure*}

We searched for periodic variability in the high-cadence KELT data prior to and during the 2009 dimming using the Lomb-Scargle (LS) periodicity search algorithm \citep{Lomb:1976, Scargle:1982} in the VARTOOLS analysis package \citep{Hartman:2012}. For the each season of KELT data, we find a 0.3214 day period with an average peak-to-peak amplitude of $\sim$2\% (see Figure \ref{figure:phased}).  This periodic signal is clearly changing in amplitude and phase over the 9 years of observations. We find the same $\sim$0.32 day periodicity in every season except the 2008-2009 observing season where the source faded. Additionally, the targeted observations by MINERVA and KUO are consistent with this periodicity. There is no color evolution over the short duration of these targeted observations. During the 2009 dimming, the $\sim$0.32 day periodicity is recovered but at a lower signal-to-noise ratio. The periodicity is likely caused by processes near the star. If the 0.32 day periodicity is the rotation period of the star, then this would imply a rotational velocity of $\sim 347\pm 57$\,\kms for a stellar radius of 2.619$^{+0.597}_{-0.304}$\rsun, compared to a breakup period of $0.39\pm0.13$ days. Given our uncertainty in the stellar radius, V1334 Tau A could be rotating near break up velocity. We explore the possibility that V1334 Tau is a rapidly rotating star close to a pole-on orientation in \S \ref{sec:interp}. Another possibility is that the $\sim$0.32 day period is the orbital period of some circumstellar material that periodically obscures the star at a semimajor axis of $\sim$2.3R$_{\star}$. 

However, the observed periodicity is persistent over the entire KELT data set, making it unlikely to be related to accretion processes. Additionally, if there is a significant amount of orbiting disk material right at the surface of the star, we would expect to see an IR excess in the SED analysis. The fourteen spectra of V1334 Tau on UT 2016 December 09 and 10 show no significant large radial velocity variability ruling out the possibility of a close stellar binary. The orbital semi-amplitude from the TRES observations is 338$\pm$56 \ms, when a photometric ephemeris with a period 0.3124 days is imposed. The velocity residuals from a circular orbit have an RMS of 433 m/s. The RMS of all the RV measurements is 1.3 \kms. Using our determined stellar mass of 1.67 \msun and a period of 0.3214 days, we get an upper mass limit of a companion to be $<$2 \mj (RMS = 433 \ms) or $<$6.2 \mj (RMS = 1.3 \kms).  The upcoming K2 campaign 13 observations of V1334 Tau should provide additional information on the nature of this short-period variability.



\subsection{Pre-2004 Dimming}
In UT 2002 December V1334 Tau was in its ``minimum'' state, at a $V$ $\sim$ 9.7 mag. The system took $\sim$280 days to return to its quiescent brightness of $\sim$9.6 mag. Unfortunately, there are no available photometric observations of V1334 Tau prior to UT 2002 December, so there is no constraint on the dimming duration. We only know that the dimming must be longer than the time between the start of the ASAS observations and the end of the observed brightening - i.e. greater than $\sim$320 days. 

\subsection{2009 Dimming}
During late February of 2009, the V1334 Tau system began to fade from its median magnitude of $V$ $\sim$ 9.59 mag. We estimate that the ingress of the dimming takes $\sim$273 days and that the event has a maximum depth of $\sim$0.12 mag ($V$ $\sim$ 9.71, the maximum depth occurred during the 2012 and 2016 observing seasons). This ingress timescale is very similar to the estimated timescale to brighten in 2002-2003. The dimming lasts through the end of the data analyzed here on UT 2016 November 07, leading to a lower limit on the dimming duration of $>$2802 days. If we assume that the two dimming events are similar, the time between events is $>$13 years. Multi-band observations from MINERVA and KUO show no $B-V$ color trend over the course of an $\sim$6 hour observing night. 

With the combined high cadence observations from KELT and WASP during the 2009 dimming, we are able to study the photometric variability during the event. If this is indeed an occultation, the additional brightening and dimming events seen during the large dimming (on a similar timescale as the initial dimming) suggests that we are observing sub-structure in the occulting body. After the ingress, the system stays at a constant brightness of $\sim$9.66 mag through the remainder of the 2009-2010 observing season and the first half of the 2010-2011 season. At JD $\sim$ 2455502, V1334 Tau begins to dim again, down to brightness of $\sim$9.69 mag. At the start of the 2011-2012 observing season, V1334 Tau is near its minimum brightness of $\sim$9.71 mag. Since the end of this additional dimming occurred during the 2011 seasonal observing gap, we can only place a upper limit of $\sim$325 days on its duration. After V1334 Tau brightens to $\sim$9.65 in 2012, it remains at a depth of $\sim$9.69 mag for the following season. Following the 2013/2014 season, V1334 Tau brightens over the course of the entire season from $\sim$9.69 mag to $\sim$9.64 mag. Following this peak in brightness, the following two seasons of ASAS-SN, KUO, and MINERVA data suggest that V1334 Tau faded back to the maximum depth of $\sim$9.71 mag. Interestingly, the timescales of these variation are similar to the estimated ingress timescale in 2009. However, our ability to draw any conclusions is limited by the seasonal observing gaps. We discuss the possibility that the observed structure of the dimming may be caused by a change in the opacity of the occulting feature in \S7. 

\subsection{Spectroscopic Analysis}
\label{sec:spec}
\begin{figure*}[!ht]
\centering
\begin{tabular}{cc}
    \includegraphics[width=0.4\linewidth]{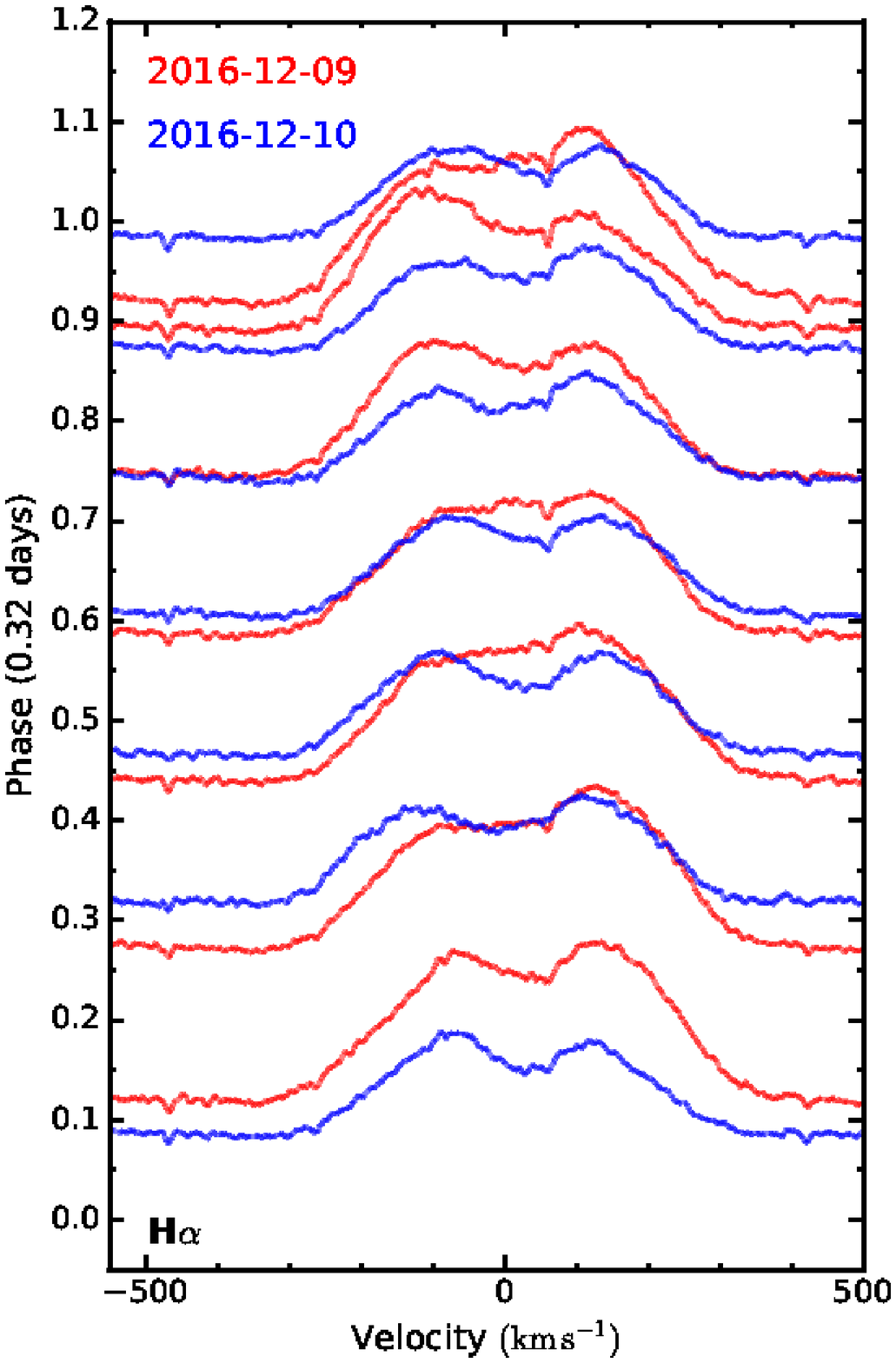} &  
    \includegraphics[width=0.4\linewidth]{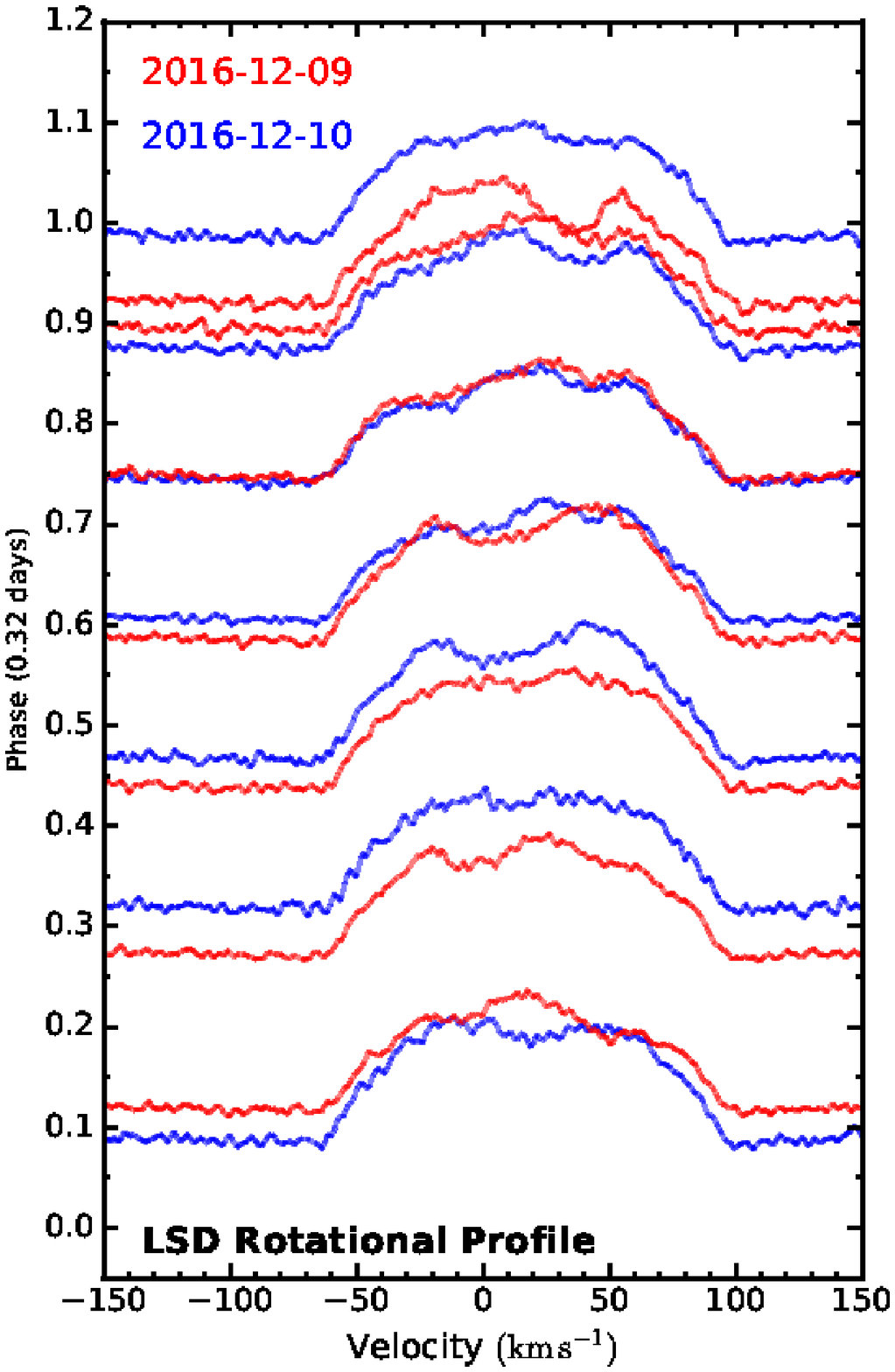}\\  
\end{tabular}
\caption{The H$\alpha$spectral line profiles of V1334 Tau show dramatic variations over the course of hours. \textbf{Left:} The double-peaked H$\alpha$ line profile for the TRES observations on UT 2016 December 09 (Red) and 10 (Blue) phased to the 0.32 day photometric modulation period (see Section~\ref{sec:oot_phot-var}). Note that the line profile changes significantly over the span of the observations. \textbf{Right:} Rotational broadening profiles derived from each TRES spectrum, derived via a least-squares deconvolution process and again phased to the 0.32 day period.}
\label{HalphaFigure}
\end{figure*}

To check for photospheric features on the stellar disk, we derived line broadening profiles using a least-squares deconvolution (LSD) of the TRES spectra (following \citealp{Zhou:2016}). Indentations and protrusions in the broadening profile map the presence of spots, faculae, and pulsation on the stellar surface. The broadening profiles show large-scale variations over the course of hours. Figure~\ref{HalphaFigure} shows the time series of the spectra gathered over 2016 December 09 and 10, phased to the 0.3214 day photometric modulation period (see Section~\ref{sec:oot_phot-var}). We find that the broadening profiles do not phase well with the photometric period, suggesting that there is no persistent spot crossing the star, and that the variability is possibly due to non-radial pulsations on the star. In addition, we find no evidence of line blending indicative of a binary stellar companion orbiting the primary star. 


We also find a double-peaked H$\alpha$ emission line, which may be an indication of an accretion disk (See Figure~\ref{HalphaFigure}). The H$\alpha$ feature has an equivalent width of $\sim 5$\AA, which could be taken to mean that the star still possesses an inner accretion disk \citep{Mohanty:2005}. The relatively high equivalent width also suggests that V1334 Tau should be reclassified as a Classical T-Tauri Star (CTTS) ($>$3\AA, \citealp{White:2003}). Given our mass estimates, the peak H$\alpha$ emission corresponds to a distance of $>$0.16\,AU from the star, assuming we are observing a Keplerian velocity for a possible inner disk. However, it is also possible that the H$\alpha$ feature could be from cooler, neutral HI that is falling onto the star or disk winds. No other Balmer lines are found in emission, nor do we find other common accretion signatures, such as the He I 5876\AA\space and He I 6670\AA\space lines in emission. We see the Ca II H and K lines with core emission, and equivalent widths of 1.5\AA\space and 2.6\AA\space respectively, perhaps indicative of the strong chromospheric activity expected for T Tauri stars. Lithium absorption at 6708\AA\space is also present, with an equivalent width of 0.25\AA, consistent with the absence of depletion expected for a solar-type star with an age of 2\,Myr \citep{Soderblom:1993}. The sodium Na I D feature has a depth and width consistent with a stellar absorption line, along with a superimposed narrow feature due to the interstellar medium. Unlike RW Aur \citep{Facchini:2016}, we do not see a time varying P Cygni profile in the Na lines, which would suggest the existence of a strong disk wind. However, we do not have a spectroscopic observation of V1334 Tau prior to the large dimming, limiting our interpretation. 

\section{Interpretation and Discussion}
\label{sec:interp}
With only two incompletely sampled dimming events, we are unable to determine whether this phenomenon is periodic in nature. However, the egress/ingress timescales from each event are consistent. In similar systems with dimming events, such as V409 Tau and DM Ori, the dimmings are believed to be periodic \citep{Rodriguez:2015, Rodriguez:2016C}. In this section we explore the possibility that the dimmings of V1334 Tau is caused by an orbiting body (e.g. a disk warp or dust trap), enhanced disk winds, hydrodynamical fluctuations of the inner disk, or a significant increase in the magnetic field flux at the surface of the star.

\subsection{Occultation by Orbiting Body}
To determine the key properties of the occulting feature, we model the observed dimmings as an occultation of the host star by a large body in Keplerian motion. The leading edge of the occulting body can be perpendicular to its direction of motion ($\theta$ = 90, sharp edge) or can be inclined (``wedge-shaped'', see Figure 9 from \citealp{Rodriguez:2015}). The estimated ingress time of the 2009 dimming, where we have the best photometric coverage, is $\sim$273 days. The ingress timescale gives us an estimate of the occulting body's transverse velocity, V$_T \sim 2R_{\star}/(T_{ingress}\sin\theta)$. For R$_{\star}$ =  2.62$\rsun$, we get a minimum velocity of $\sim$155.1 \ms. The full duration of the 2009 dimming is $>$2803 days. Using our estimated transverse velocity and the duration timescale, we estimate that the occulting body must be $>$0.25 AU in width. If we assume that the occulting body is in Keplerian motion, the ingress timescale also provides an estimate of the orbital semi-major axis (see Equation 1 from \citealp{Rodriguez:2016C}). However, using these estimated parameters, the occulting body would be unrealistically far from the host star (at $\sim$62,000 AU), ruling this interpretation out. 

Unfortunately, there is no measurement of the size of the circumstellar disk around V1334 Tau A. We know from \citet{Kohler:1998} and \citet{Daemgen:2015} that there are two nearby companions to V1334 Tau A, V1334 Tau B (1$\farcs$631) and C (0$\farcs$106). Therefore, we can constrain the occulting feature to be within the projected separation of the B component, 1$\farcs$631 (215 AU for d = 132 pc). Assuming the occulting feature is at the projected separation of V1334 Tau B, it would require a transverse velocity of $\sim$2.6\kms\space and be $\sim$4.2 AU in width. Also, constraining the occulter to 219 AU would suggest a wedge-angle of $<$3.4$^{\circ}$. From our analysis of the TRES spectra, we estimate that the peak emission of the double-peaked H$\alpha$ emission corresponds to a distance of $>$0.16 AU. Assuming the occulting feature is associated with an inner disk and at a semi-major axis of 0.16 AU would imply that it is moving at $\sim$98 \kms and have a leading edge angle of 0.06$^{\circ}$. In this scenario, the duration of the occultation requires multiple orbits ($\ge$160) of the same material across our line-of-sight to cause the dimming event that began in 2009. This could be explained if the entire inner disk has precessed into our line of sight but this does not seem plausible. We can also place the occulting body at the distance of the C component (14 AU for d = 132 pc). This would suggest a transverse velocity of $\sim$10.3\kms\space, a $\sim$16.7 AU wide occulter, and a leading edge angle of 0.86$^{\circ}$. The width would be $\sim$19\% of the entire orbit at 14 AU, assuming Keplerian motion. At 14 AU, an orbital period would be $\sim$40 years. It is possible that the close-in nearby companion is associated with the observed dimming events.

Fortunately, we observed two separate dimmings, one pre-2004 and one beginning in 2009. Since the egress from the pre-2004 dimming and the ingress of the 2009 dimming are similar in duration, it is plausible that we are seeing a periodic occultation with a period of $>$13 years. For a stellar mass of 1.67 $\msun$, this period would suggest that the occulting feature is located $\ge$6.6 AU from the host star. Using the same wedge-shaped model, the leading edge of the occulting body would have a highly inclined leading edge angle of $\sim$0.6$^{\circ}$, almost parallel to its direction of motion. This would also suggest that the occulter is moving at $\sim$15\kms and be $\ge$24 AU in width, which is $>$59\% of the entire orbit. 

Observations from the Atacama Large Millimeter/submillimeter Array (ALMA) and the Hubble Space Telescope (HST) have shown that transitional disks can have large asymmetries \citep{Perez:2014, Stark:2014}. Specifically, HST observations of MP Mus have shown that the eastern side of its disk is three times brighter then its western side \citep{Schneider:2014}. The young star Oph IRS 48 has a large asymmetric dust structure that extends across almost one third of the entire disk ring \citep{vandermarel:2013}. Therefore, if V1334 Tau has an edge-on, asymmetric disk, the large dimming could be caused by a dust trap in the disk crossing our line-of-sight leading to a periodic dimming. Future millimeter mapping of the V1334 Tau system should determine the plausibility of this interpretation. However, no detectable IR excess out to 20 microns would suggest a very low mass, nearly transparent circumstellar disk, making this interpretation unlikely.   

\subsection{Inner Disk Obscuration}
An alternate explanation for the UXor phenomena is hydrodynamical fluctuations of the inner disk rim. \citet{Dullemond:2003} determined that the inner disk rim would have a scale height of $\sim$0.2 R$_{rim}$ and fluctuate by $\pm$ $\sim$0.1 R$_{rim}$. Using 0.16 AU as an estimate of the inner disk radius, this would suggest a maximum fluctuation of the inner rim height to be $\sim$0.048 AU. In addition to the hydrodynamical fluctuation, the inner disk of a YSO can be warped and perturbed by a misaligned close companion \citep{Facchini:2013, Facchini:2014}. Without a constraint on the inner disk inclination, we are unable to test the feasibility of this interpretation. However, this scenario has been proposed as the cause of the observed dimming events of RW Aur \citep{Facchini:2016}. 

In any case, the variability during the dimming suggests an opacity change in the occulting feature. Another interpretation for the extended dimming is that a screen of hot dust can be created in the inner disk and pushed across our line-of-sight by high disk winds \citep{Petrov:2015}. However, we find no evidence for high or enhanced disk winds in the TRES spectra, making this scenario unlikely. Future millimeter mapping of the disk around V1334 Tau and radial velocity monitoring for a short period companion will better constrain these scenarios. 

\subsection{Increased Magnetic Field Flux}
Another possible explanation of the dimming events seen in V1334 Tau is they are a result of significant increase in the magnetic field flux at the surface of the star. According to the MIST stellar evolution models, the interior structure of V1334 Tau is predicted to be fully convective. Current magneto-hydrodynamic stellar structure models of fully convective dwarfs suggest that these stars are able to support stable ($>$ 1000 days), large-scale dipolar fields ($>$ 1 kilogauss) despite rotating rapidly, resulting in the formation of large groupings of dark polar starspots blanketing a significant fraction of the stellar photosphere \citep{Yadav:2015}. The formation of long-lived polar spots may result in multi-year dimming events similar to the ones seen here, particularly if we are viewing this star nearly pole-on. The high amount of H$\alpha$ emission, and the excess NUV and FUV GALEX flux observed in the star's SED, suggests that V1334 Tau is magnetically active. Future observations should clarify if the dimming we observe is connected to magnetic activity. This includes stellar color information in and out of dimming events, mapping the stellar magnetic fields using Zeeman-Doppler Imaging, and modeling the upcoming K2 high precision light curves.


\subsection{Similar Systems}
There are only a dozen or so objects known to show this type of phenomena. The duration of the V1334 Tau dimming events strongly resemble the long term dimming of the archetype AA Tau \citep{Bouvier:2013}. Although similar in duration, the dimming event in AA Tau shows a much larger depth of $\sim$2 mag. Other systems like V409 Tau, DM Ori, and RW Aur have shown similar dimmings (again at much larger depths), and are interpreted as either a periodically obscuring feature in the circumstellar disk, high disk winds, or inner disk instabilities \citep{Rodriguez:2013, Rodriguez:2015, Rodriguez:2016A, Rodriguez:2016C, Petrov:2015, Facchini:2016}. It is possible that the dimming mechanisms for all these systems are related.

\section{Summary and Conclusions}
The complex circumstellar environment of YSOs can directly influence planetary formation and architecture. Unfortunately, it is not clear how each process or feature of a circumstellar disk (accretion, winds, warps, and companions) affects the planet forming process. From studying these processes for a large sample of YSOs at different ages, we may gain insight into the role each feature has on planet formation.

Here, we have discovered two dimming events in the WLTS V1334 Tau. The first fading occurred prior to 2004, where we observe a $\sim$0.1 mag dimming over $\sim$300 days. The second event started in 2009 and has yet to end. The photometric observations show a coherent dimming of $\sim$0.12 mag, an ingress dimming timescale of $\sim$273 days, and a duration of $>$2803 days. From applying a simple model where the dimming is caused by an occultation of V1334 Tau A by a large semi-opaque body that has an inclined leading edge, we determined that the occulting body would have to be moving at $\sim$14.2 \kms, have a semi-major axis of $\sim$6.56 AU, and be $\sim$24.3 AU in width. Assuming a circular orbit, this would suggest that the occulting feature is quite large, as the width is $\sim$59\% of its entire orbit. This could be explained by an asymmetric disk or dust trap occulting the star in a nearly edge-on orientated disk. Although we see no IR excess, the double-peaked H$\alpha$ line seen in the TRES spectra suggests the presence of orbiting material. It is possible that the occulting body may be a warp or perturbation in the surrounding disk, but this would require a close to edge-on geometry. It is possible that we are seeing a fluctuation in the inner disk structure or that winds have pushed material from the disk plane across our line of sight \citep{Natta:2000, Nelson:2000, Petrov:2015}. However, we find no evidence for high disk winds in our TRES spectra. We also find a $\sim$0.32 day periodicity in the KELT photometry that persists through the 2009 dimming. The specific cause of this periodicity is unclear but it is consistent with orbiting material only a few stellar radii from the star. 

Understanding the observed extended dimmings of V1334 Tau over the past $\sim$15 years requires the continued photometric monitoring of the system. As a start, V1334 Tau will be observed in the upcoming K2 campaign 13 (spring 2017). Simultaneous multi-band photometric and spectroscopic observations leading up to and coinciding with the K2 observations may provide new information to distinguish between potential dimming mechanisms. Future millimeter mapping of the V1334 Tau system should detect and measure the inclination of the possible circumstellar disk. Since V1334 Tau is a visual binary, millimeter mapping may also show evidence of a stellar encounter, similar to what has been seen in RW Aur \citep{Cabrit:2006}. Upcoming large surveys like the Transiting Exoplanet Survey Satellite (TESS, \citealp{Ricker:2015}), the Large Synoptic Survey Telescope (LSST, \citealp{LSST:2009}), and the PLAnetary Transits and Oscillations of stars (PLATO, \citealp{Catala:2010}) mission should increase our sample of these unique systems, allowing us to statistically address the variety in the depth and timescales of these dimming events. 

\acknowledgments
Early work on KELT-North was supported by NASA Grant NNG04GO70G. J.A.P. and K.G.S. acknowledge support from the Vanderbilt Office of the Provost through the Vanderbilt Initiative in Data-intensive Astrophysics. This work has made use of NASA's Astrophysics Data System and the SIMBAD database operated at CDS, Strasbourg, France.

Work performed by J.E.R. was supported by the Harvard Future Faculty Leaders Postdoctoral fellowship. Work performed by P.A.C. was supported by NASA grant NNX13AI46G. Work by K.G.S. was supported by NSF PAARE grant AST-1358862. B.J.S. is supported by NASA through Hubble Fellowship grant HF-51348.001 awarded by the Space Telescope Science Institute, which is operated by the Association of Universities for Research in Astronomy, Inc., for NASA, under contract NAS 5-26555.  CSK and KZS are supported by NSF grants AST-1515876 and AST-1515927. TW-SH is supported by the DOE Computational Science Graduate Fellowship, grant number DE-FG02-97ER25308. P.C. is supported by the NASA grant NNX13AI46G. Work by D.J.S. and B.S.G. was partially supported by NSF CAREER Grant AST-1056524.

Development of ASAS-SN has been supported by NSF grant AST-0908816 and CCAPP at the Ohio State University.  ASAS-SN is supported by NSF grant AST-1515927, the Center for Cosmology and AstroParticle Physics (CCAPP) at OSU, the Mt. Cuba Astronomical Foundation, George Skestos, and the Robert Martin Ayers Sciences Fund.

This paper makes use of data from the first public release of the WASP data \citep{Butters:2010} as provided by the WASP consortium and services at the NASA Exoplanet Archive, which is operated by the California Institute of Technology, under contract with the National Aeronautics and Space Administration under the Exoplanet Exploration Program.

MINERVA is a collaboration among the Harvard-Smithsonian Center for Astrophysics, The Pennsylvania State University, the University of Montana, and the University of New South Wales. MINERVA is made possible by generous contributions from its collaborating institutions and Mt. Cuba Astronomical Foundation, The David \& Lucile Packard Foundation, National Aeronautics and Space Administration (EPSCOR grant NNX13AM97A), The Australian Research Council (LIEF grant LE140100050), and the National Science Foundation (grants 1516242 and 1608203). Any opinions, findings, and conclusions or recommendations expressed are those of the author and do not necessarily reflect the views of the National Science Foundation. This work was partially supported by funding from the Center for Exoplanets and Habitable Worlds, which is supported by the Pennsylvania State University, the Eberly College of Science, and the Pennsylvania Space Grant Consortium.

\bibliographystyle{apj}

\bibliography{V1334Tau}

\begin{thebibliography}{}
\expandafter\ifx\csname natexlab\endcsname\relax\def\natexlab#1{#1}\fi

\bibitem[{{Bensby} {et~al.}(2003){Bensby}, {Feltzing}, \&
  {Lundstr{\"o}m}}]{Bensby:2003}
{Bensby}, T., {Feltzing}, S., \& {Lundstr{\"o}m}, I. 2003, \aap, 410, 527

\bibitem[{{Bertout:} \& {Genova}(2006)}]{Bertout:2006}
{Bertout:}, C., \& {Genova}, F. 2006, \aap, 460, 499

\bibitem[{{Bouvier} {et~al.}(2013){Bouvier}, {Grankin}, {Ellerbroek}, {Bouy},
  \& {Barrado}}]{Bouvier:2013}
{Bouvier}, J., {Grankin}, K., {Ellerbroek}, L.~E., {Bouy}, H., \& {Barrado}, D.
  2013, \aap, 557, A77

\bibitem[{{Bouvier} {et~al.}(2007){Bouvier}, {Alencar}, {Boutelier},
  {Dougados}, {Balog}, {Grankin}, {Hodgkin}, {Ibrahimov}, {Kun}, {Magakian}, \&
  {Pinte}}]{Bouvier:2007}
{Bouvier}, J., {Alencar}, S.~H.~P., {Boutelier}, T., {et~al.} 2007, \aap, 463,
  1017

\bibitem[{{Brice{\~n}o} {et~al.}(2002){Brice{\~n}o}, {Luhman}, {Hartmann},
  {Stauffer}, \& {Kirkpatrick}}]{Briceno:2002}
{Brice{\~n}o}, C., {Luhman}, K.~L., {Hartmann}, L., {Stauffer}, J.~R., \&
  {Kirkpatrick}, J.~D. 2002, \apj, 580, 317

\bibitem[{{Brown} {et~al.}(2013){Brown}, {Baliber}, {Bianco}, {Bowman},
  {Burleson}, {Conway}, {Crellin}, {Depagne}, {De Vera}, {Dilday}, {Dragomir},
  {Dubberley}, {Eastman}, {Elphick}, {Falarski}, {Foale}, {Ford}, {Fulton},
  {Garza}, {Gomez}, {Graham}, {Greene}, {Haldeman}, {Hawkins}, {Haworth},
  {Haynes}, {Hidas}, {Hjelstrom}, {Howell}, {Hygelund}, {Lister}, {Lobdill},
  {Martinez}, {Mullins}, {Norbury}, {Parrent}, {Paulson}, {Petry}, {Pickles},
  {Posner}, {Rosing}, {Ross}, {Sand}, {Saunders}, {Shobbrook}, {Shporer},
  {Street}, {Thomas}, {Tsapras}, {Tufts}, {Valenti}, {Vander Horst}, {Walker},
  {White}, \& {Willis}}]{Brown:2013}
{Brown}, T.~M., {Baliber}, N., {Bianco}, F.~B., {et~al.} 2013, \pasp, 125, 1031

\bibitem[{{Butters} {et~al.}(2010){Butters}, {West}, {Anderson}, {Collier
  Cameron}, {Clarkson}, {Enoch}, {Haswell}, {Hellier}, {Horne}, {Joshi},
  {Kane}, {Lister}, {Maxted}, {Parley}, {Pollacco}, {Smalley}, {Street},
  {Todd}, {Wheatley}, \& {Wilson}}]{Butters:2010}
{Butters}, O.~W., {West}, R.~G., {Anderson}, D.~R., {et~al.} 2010, \aap, 520,
  L10

\bibitem[{{Cabrit} {et~al.}(2006){Cabrit}, {Pety}, {Pesenti}, \&
  {Dougados}}]{Cabrit:2006}
{Cabrit}, S., {Pety}, J., {Pesenti}, N., \& {Dougados}, C. 2006, \aap, 452, 897

\bibitem[{{Carroll} {et~al.}(1991){Carroll}, {Guinan}, {McCook}, \&
  {Donahue}}]{Carroll:1991}
{Carroll}, S.~M., {Guinan}, E.~F., {McCook}, G.~P., \& {Donahue}, R.~A. 1991,
  \apj, 367, 278

\bibitem[{{Catala} {et~al.}(2010){Catala}, {Arentoft}, {Fridlund}, {Lindberg},
  {Mas-Hesse}, {Micela}, {Pollacco}, {Poretti}, {Rauer}, {Roxburgh}, {Stankov},
  \& {Udry}}]{Catala:2010}
{Catala}, C., {Arentoft}, T., {Fridlund}, M., {et~al.} 2010, in Astronomical
  Society of the Pacific Conference Series, Vol. 430, Pathways Towards
  Habitable Planets, ed. V.~{Coud{\'e} du Foresto}, D.~M. {Gelino}, \&
  I.~{Ribas}, 260

\bibitem[{{Choi} {et~al.}(2016){Choi}, {Dotter}, {Conroy}, {Cantiello},
  {Paxton}, \& {Johnson}}]{Choi:2016}
{Choi}, J., {Dotter}, A., {Conroy}, C., {et~al.} 2016, ArXiv e-prints,
  arXiv:1604.08592

\bibitem[{{Co{\c s}kuno{\v g}lu} {et~al.}(2011){Co{\c s}kuno{\v g}lu}, {Ak},
  {Bilir}, {Karaali}, {Yaz}, {Gilmore}, {Seabroke}, {Bienaym{\'e}},
  {Bland-Hawthorn}, {Campbell}, {Freeman}, {Gibson}, {Grebel}, {Munari},
  {Navarro}, {Parker}, {Siebert}, {Siviero}, {Steinmetz}, {Watson}, {Wyse}, \&
  {Zwitter}}]{Coskunoglu:2011}
{Co{\c s}kuno{\v g}lu}, B., {Ak}, S., {Bilir}, S., {et~al.} 2011, \mnras, 412,
  1237

\bibitem[{{Cody} {et~al.}(2014){Cody}, {Stauffer}, {Baglin}, {Micela},
  {Rebull}, {Flaccomio}, {Morales-Calder{\'o}n}, {Aigrain}, {Bouvier},
  {Hillenbrand}, {Gutermuth}, {Song}, {Turner}, {Alencar}, {Zwintz},
  {Plavchan}, {Carpenter}, {Findeisen}, {Carey}, {Terebey}, {Hartmann},
  {Calvet}, {Teixeira}, {Vrba}, {Wolk}, {Covey}, {Poppenhaeger}, {G{\"u}nther},
  {Forbrich}, {Whitney}, {Affer}, {Herbst}, {Hora}, {Barrado}, {Holtzman},
  {Marchis}, {Wood}, {Medeiros Guimar{\~a}es}, {Lillo Box}, {Gillen},
  {McQuillan}, {Espaillat}, {Allen}, {D'Alessio}, \& {Favata}}]{Cody:2014}
{Cody}, A.~M., {Stauffer}, J., {Baglin}, A., {et~al.} 2014, \aj, 147, 82

\bibitem[{{Cutri} {et~al.}(2012){Cutri}, {Wright}, T., \& {at
  al.}}]{Cutri:2012}
{Cutri}, R.~M., {Wright}, E.~L., T., C., \& {at al.} 2012, VizieR Online Data
  Catalog, 2311, 0

\bibitem[{{Cutri} {et~al.}(2003){Cutri}, {Skrutskie}, {van Dyk}, {Beichman},
  {Carpenter}, {Chester}, {Cambresy}, {Evans}, {Fowler}, {Gizis}, {Howard},
  {Huchra}, {Jarrett}, {Kopan}, {Kirkpatrick}, {Light}, {Marsh}, {McCallon},
  {Schneider}, {Stiening}, {Sykes}, {Weinberg}, {Wheaton}, {Wheelock}, \&
  {Zacarias}}]{Cutri:2003}
{Cutri}, R.~M., {Skrutskie}, M.~F., {van Dyk}, S., {et~al.} 2003, VizieR Online
  Data Catalog, 2246, 0

\bibitem[{{Daemgen} {et~al.}(2015){Daemgen}, {Bonavita}, {Jayawardhana},
  {Lafreni{\`e}re}, \& {Janson}}]{Daemgen:2015}
{Daemgen}, S., {Bonavita}, M., {Jayawardhana}, R., {Lafreni{\`e}re}, D., \&
  {Janson}, M. 2015, \apj, 799, 155

\bibitem[{{Dotter} {et~al.}(2008){Dotter}, {Chaboyer}, {Jevremovi{\'c}},
  {Kostov}, {Baron}, \& {Ferguson}}]{Dotter:2008}
{Dotter}, A., {Chaboyer}, B., {Jevremovi{\'c}}, D., {et~al.} 2008, \apjs, 178,
  89

\bibitem[{{Dullemond} {et~al.}(2003){Dullemond}, {van den Ancker}, {Acke}, \&
  {van Boekel}}]{Dullemond:2003}
{Dullemond}, C.~P., {van den Ancker}, M.~E., {Acke}, B., \& {van Boekel}, R.
  2003, \apjl, 594, L47

\bibitem[{{Facchini} {et~al.}(2013){Facchini}, {Lodato}, \&
  {Price}}]{Facchini:2013}
{Facchini}, S., {Lodato}, G., \& {Price}, D.~J. 2013, \mnras, 433, 2142

\bibitem[{{Facchini} {et~al.}(2016){Facchini}, {Manara}, {Schneider}, {Clarke},
  {Bouvier}, {Rosotti}, {Booth}, \& {Haworth}}]{Facchini:2016}
{Facchini}, S., {Manara}, C.~F., {Schneider}, P.~C., {et~al.} 2016, ArXiv
  e-prints, arXiv:1610.08303

\bibitem[{{Facchini} {et~al.}(2014){Facchini}, {Ricci}, \&
  {Lodato}}]{Facchini:2014}
{Facchini}, S., {Ricci}, L., \& {Lodato}, G. 2014, \mnras, 442, 3700

\bibitem[{{Feroz} {et~al.}(2009){Feroz}, {Hobson}, \& {Bridges}}]{Feroz:2009}
{Feroz}, F., {Hobson}, M.~P., \& {Bridges}, M. 2009, \mnras, 398, 1601

\bibitem[{{G{\'o}mez de Castro} {et~al.}(2015){G{\'o}mez de Castro},
  {Lopez-Santiago}, {L{\'o}pez-Mart{\'{\i}}nez}, {S{\'a}nchez}, {Sestito}, {de
  Castro}, {Cornide}, \& {Ya{\~n}ez Gestoso}}]{Gomez:2015}
{G{\'o}mez de Castro}, A.~I., {Lopez-Santiago}, J.,
  {L{\'o}pez-Mart{\'{\i}}nez}, F., {et~al.} 2015, \apjs, 216, 26

\bibitem[{{Grady} {et~al.}(2000){Grady}, {Sitko}, {Russell}, {Lynch}, {Hanner},
  {Perez}, {Bjorkman}, \& {de Winter}}]{Grady:2000}
{Grady}, C.~A., {Sitko}, M.~L., {Russell}, R.~W., {et~al.} 2000, Protostars and
  Planets IV, 613

\bibitem[{{Grinin}(1988)}]{Grinin:1988}
{Grinin}, V.~P. 1988, Soviet Astronomy Letters, 14, 27

\bibitem[{{Grinin} {et~al.}(1998){Grinin}, {Rostopchina}, \&
  {Shakhovskoi}}]{Grinin:1998}
{Grinin}, V.~P., {Rostopchina}, A.~N., \& {Shakhovskoi}, D.~N. 1998, Astronomy
  Letters, 24, 802

\bibitem[{{Hartman}(2012)}]{Hartman:2012}
{Hartman}, J. 2012, {VARTOOLS: Light Curve Analysis Program}, Astrophysics
  Source Code Library, ascl:1208.016

\bibitem[{{Henden} {et~al.}(2016){Henden}, {Templeton}, {Terrell}, {Smith},
  {Levine}, \& {Welch}}]{Henden:2016}
{Henden}, A.~A., {Templeton}, M., {Terrell}, D., {et~al.} 2016, VizieR Online
  Data Catalog, 2336

\bibitem[{{Herbst} {et~al.}(1994){Herbst}, {Herbst}, {Grossman}, \&
  {Weinstein}}]{Herbst:1994}
{Herbst}, W., {Herbst}, D.~K., {Grossman}, E.~J., \& {Weinstein}, D. 1994, \aj,
  108, 1906

\bibitem[{{H{\o}g} {et~al.}(2000){H{\o}g}, {Fabricius}, {Makarov}, {Urban},
  {Corbin}, {Wycoff}, {Bastian}, {Schwekendiek}, \& {Wicenec}}]{Hog:2000}
{H{\o}g}, E., {Fabricius}, C., {Makarov}, V.~V., {et~al.} 2000, \aap, 355, L27

\bibitem[{{Kloppenborg} {et~al.}(2010){Kloppenborg}, {Stencel}, {Monnier},
  {Schaefer}, {Zhao}, {Baron}, {McAlister}, {ten Brummelaar}, {Che},
  {Farrington}, {Pedretti}, {Sallave-Goldfinger}, {Sturmann}, {Sturmann},
  {Thureau}, {Turner}, \& {Carroll}}]{Kloppenborg:2010}
{Kloppenborg}, B., {Stencel}, R., {Monnier}, J.~D., {et~al.} 2010, \nat, 464,
  870

\bibitem[{{Kohler} \& {Leinert}(1998)}]{Kohler:1998}
{Kohler}, R., \& {Leinert}, C. 1998, \aap, 331, 977

\bibitem[{{Lomb}(1976)}]{Lomb:1976}
{Lomb}, N.~R. 1976, \apss, 39, 447

\bibitem[{{LSST Science Collaboration} {et~al.}(2009){LSST Science
  Collaboration}, {Abell}, {Allison}, {Anderson}, {Andrew}, {Angel}, {Armus},
  {Arnett}, {Asztalos}, {Axelrod}, \& et~al.}]{LSST:2009}
{LSST Science Collaboration}, {Abell}, P.~A., {Allison}, J., {et~al.} 2009,
  ArXiv e-prints, arXiv:0912.0201

\bibitem[{{Luhman} {et~al.}(2003){Luhman}, {Stauffer}, {Muench}, {Rieke},
  {Lada}, {Bouvier}, \& {Lada}}]{Luhman:2003}
{Luhman}, K.~L., {Stauffer}, J.~R., {Muench}, A.~A., {et~al.} 2003, \apj, 593,
  1093

\bibitem[{{Mohanty} {et~al.}(2005){Mohanty}, {Jayawardhana}, \&
  {Basri}}]{Mohanty:2005}
{Mohanty}, S., {Jayawardhana}, R., \& {Basri}, G. 2005, \apj, 626, 498

\bibitem[{{Natta} \& {Whitney}(2000)}]{Natta:2000}
{Natta}, A., \& {Whitney}, B.~A. 2000, \aap, 364, 633

\bibitem[{{Nelson} {et~al.}(2000){Nelson}, {Benz}, \&
  {Ruzmaikina}}]{Nelson:2000}
{Nelson}, A.~F., {Benz}, W., \& {Ruzmaikina}, T.~V. 2000, \apj, 529, 357

\bibitem[{{Nguyen} {et~al.}(2012){Nguyen}, {Brandeker}, {van Kerkwijk}, \&
  {Jayawardhana}}]{Nguyen:2012}
{Nguyen}, D.~C., {Brandeker}, A., {van Kerkwijk}, M.~H., \& {Jayawardhana}, R.
  2012, \apj, 745, 119

\bibitem[{{Pepper} {et~al.}(2012){Pepper}, {Kuhn}, {Siverd}, {James}, \&
  {Stassun}}]{Pepper:2012}
{Pepper}, J., {Kuhn}, R.~B., {Siverd}, R., {James}, D., \& {Stassun}, K. 2012,
  \pasp, 124, 230

\bibitem[{{Pepper} {et~al.}(2007){Pepper}, {Pogge}, {DePoy}, {Marshall},
  {Stanek}, {Stutz}, {Poindexter}, {Siverd}, {O'Brien}, {Trueblood}, \&
  {Trueblood}}]{Pepper:2007}
{Pepper}, J., {Pogge}, R.~W., {DePoy}, D.~L., {et~al.} 2007, \pasp, 119, 923

\bibitem[{{P{\'e}rez} {et~al.}(2014){P{\'e}rez}, {Isella}, {Carpenter}, \&
  {Chandler}}]{Perez:2014}
{P{\'e}rez}, L.~M., {Isella}, A., {Carpenter}, J.~M., \& {Chandler}, C.~J.
  2014, \apjl, 783, L13

\bibitem[{{Petrov} {et~al.}(2015){Petrov}, {Gahm}, {Djupvik}, {Babina},
  {Artemenko}, \& {Grankin}}]{Petrov:2015}
{Petrov}, P.~P., {Gahm}, G.~F., {Djupvik}, A.~A., {et~al.} 2015, \aap, 577, A73

\bibitem[{{Petrov} \& {Kozack}(2007)}]{Petrov:2007}
{Petrov}, P.~P., \& {Kozack}, B.~S. 2007, Astronomy Reports, 51, 500

\bibitem[{{Pojmanski}(1997)}]{Pojamanski:1997}
{Pojmanski}, G. 1997, Acta Astronomica, 47, 467

\bibitem[{{Ricker} {et~al.}(2015){Ricker}, {Winn}, {Vanderspek}, {Latham},
  {Bakos}, {Bean}, {Berta-Thompson}, {Brown}, {Buchhave}, {Butler}, {Butler},
  {Chaplin}, {Charbonneau}, {Christensen-Dalsgaard}, {Clampin}, {Deming},
  {Doty}, {De Lee}, {Dressing}, {Dunham}, {Endl}, {Fressin}, {Ge}, {Henning},
  {Holman}, {Howard}, {Ida}, {Jenkins}, {Jernigan}, {Johnson}, {Kaltenegger},
  {Kawai}, {Kjeldsen}, {Laughlin}, {Levine}, {Lin}, {Lissauer}, {MacQueen},
  {Marcy}, {McCullough}, {Morton}, {Narita}, {Paegert}, {Palle}, {Pepe},
  {Pepper}, {Quirrenbach}, {Rinehart}, {Sasselov}, {Sato}, {Seager},
  {Sozzetti}, {Stassun}, {Sullivan}, {Szentgyorgyi}, {Torres}, {Udry}, \&
  {Villasenor}}]{Ricker:2015}
{Ricker}, G.~R., {Winn}, J.~N., {Vanderspek}, R., {et~al.} 2015, Journal of
  Astronomical Telescopes, Instruments, and Systems, 1, 014003

\bibitem[{{Rodriguez} {et~al.}(2016{\natexlab{a}}){Rodriguez}, {Pepper}, \&
  {Stassun}}]{Rodriguez:DESK}
{Rodriguez}, J.~E., {Pepper}, J., \& {Stassun}, K.~G. 2016{\natexlab{a}}, in
  IAU Symposium, Vol. 314, Young Stars \& Planets Near the Sun, ed. J.~H.
  {Kastner}, B.~{Stelzer}, \& S.~A. {Metchev}, 167--170

\bibitem[{{Rodriguez} {et~al.}(2013){Rodriguez}, {Pepper}, {Stassun}, {Siverd},
  {Cargile}, {Beatty}, \& {Gaudi}}]{Rodriguez:2013}
{Rodriguez}, J.~E., {Pepper}, J., {Stassun}, K.~G., {et~al.} 2013, \aj, 146,
  112

\bibitem[{{Rodriguez} {et~al.}(2015){Rodriguez}, {Pepper}, {Stassun}, {Siverd},
  {Cargile}, {Weintraub}, {Beatty}, {Gaudi}, {Mamajek}, \&
  {Sanchez}}]{Rodriguez:2015}
---. 2015, The Astronomical Journal, 150, 32

\bibitem[{{Rodriguez} {et~al.}(2016{\natexlab{b}}){Rodriguez}, {Stassun},
  {Lund}, {Siverd}, {Pepper}, {Tang}, {Kafka}, {Gaudi}, {Conroy}, {Beatty},
  {Stevens}, {Shappee}, \& {Kochanek}}]{Rodriguez:2016B}
{Rodriguez}, J.~E., {Stassun}, K.~G., {Lund}, M.~B., {et~al.}
  2016{\natexlab{b}}, \aj, 151, 123

\bibitem[{{Rodriguez} {et~al.}(2016{\natexlab{c}}){Rodriguez}, {Stassun},
  {Cargile}, {Shappee}, {Siverd}, {Pepper}, {Lund}, {Kochanek}, {James},
  {Kuhn}, {Beatty}, {Gaudi}, {Weintraub}, {Stanek}, {Holoien}, {Prieto},
  {Feldman}, \& {Espaillat}}]{Rodriguez:2016C}
{Rodriguez}, J.~E., {Stassun}, K.~G., {Cargile}, P., {et~al.}
  2016{\natexlab{c}}, \apj, 831, 74

\bibitem[{{Rodriguez} {et~al.}(2016{\natexlab{d}}){Rodriguez}, {Reed},
  {Siverd}, {Pepper}, {Stassun}, {Gaudi}, {Weintraub}, {Beatty}, {Lund}, \&
  {Stevens}}]{Rodriguez:2016A}
{Rodriguez}, J.~E., {Reed}, P.~A., {Siverd}, R.~J., {et~al.}
  2016{\natexlab{d}}, \aj, 151, 29

\bibitem[{{Scargle}(1982)}]{Scargle:1982}
{Scargle}, J.~D. 1982, \apj, 263, 835

\bibitem[{{Schneider} {et~al.}(2014){Schneider}, {Grady}, {Hines}, {Stark},
  {Debes}, {Carson}, {Kuchner}, {Perrin}, {Weinberger}, {Wisniewski},
  {Silverstone}, {Jang-Condell}, {Henning}, {Woodgate}, {Serabyn},
  {Moro-Martin}, {Tamura}, {Hinz}, \& {Rodigas}}]{Schneider:2014}
{Schneider}, G., {Grady}, C.~A., {Hines}, D.~C., {et~al.} 2014, \aj, 148, 59

\bibitem[{{Shappee} {et~al.}(2014){Shappee}, {Prieto}, {Grupe}, {Kochanek},
  {Stanek}, {De Rosa}, {Mathur}, {Zu}, {Peterson}, {Pogge}, {Komossa}, {Im},
  {Jencson}, {Holoien}, {Basu}, {Beacom}, {Szczygie{\l}}, {Brimacombe},
  {Adams}, {Campillay}, {Choi}, {Contreras}, {Dietrich}, {Dubberley},
  {Elphick}, {Foale}, {Giustini}, {Gonzalez}, {Hawkins}, {Howell}, {Hsiao},
  {Koss}, {Leighly}, {Morrell}, {Mudd}, {Mullins}, {Nugent}, {Parrent},
  {Phillips}, {Pojmanski}, {Rosing}, {Ross}, {Sand}, {Terndrup}, {Valenti},
  {Walker}, \& {Yoon}}]{Shappee:2014}
{Shappee}, B.~J., {Prieto}, J.~L., {Grupe}, D., {et~al.} 2014, \apj, 788, 48

\bibitem[{{Siverd} {et~al.}(2012){Siverd}, {Beatty}, {Pepper}, {Eastman},
  {Collins}, {Bieryla}, {Latham}, {Buchhave}, {Jensen}, {Crepp}, {Street},
  {Stassun}, {Gaudi}, {Berlind}, {Calkins}, {DePoy}, {Esquerdo}, {Fulton},
  {F{\H u}r{\'e}sz}, {Geary}, {Gould}, {Hebb}, {Kielkopf}, {Marshall}, {Pogge},
  {Stanek}, {Stefanik}, {Szentgyorgyi}, {Trueblood}, {Trueblood}, {Stutz}, \&
  {van Saders}}]{Siverd:2012}
{Siverd}, R.~J., {Beatty}, T.~G., {Pepper}, J., {et~al.} 2012, \apj, 761, 123

\bibitem[{{Soderblom} {et~al.}(1993){Soderblom}, {Jones}, {Balachandran},
  {Stauffer}, {Duncan}, {Fedele}, \& {Hudon}}]{Soderblom:1993}
{Soderblom}, D.~R., {Jones}, B.~F., {Balachandran}, S., {et~al.} 1993, \aj,
  106, 1059

\bibitem[{{Stark} {et~al.}(2014){Stark}, {Schneider}, {Weinberger}, {Debes},
  {Grady}, {Jang-Condell}, \& {Kuchner}}]{Stark:2014}
{Stark}, C.~C., {Schneider}, G., {Weinberger}, A.~J., {et~al.} 2014, \apj, 789,
  58

\bibitem[{{Stassun} \& {Wood}(1999)}]{Stassun:1999}
{Stassun}, K., \& {Wood}, K. 1999, \apj, 510, 892

\bibitem[{{Swift} {et~al.}(2015){Swift}, {Bottom}, {Johnson}, {Wright},
  {McCrady}, {Wittenmyer}, {Plavchan}, {Riddle}, {Muirhead}, {Herzig}, {Myles},
  {Blake}, {Eastman}, {Beatty}, {Barnes}, {Gibson}, {Lin}, {Zhao}, {Gardner},
  {Falco}, {Criswell}, {Nava}, {Robinson}, {Sliski}, {Hedrick}, {Ivarsen},
  {Hjelstrom}, {de Vera}, \& {Szentgyorgyi}}]{swift:2015}
{Swift}, J.~J., {Bottom}, M., {Johnson}, J.~A., {et~al.} 2015, Journal of
  Astronomical Telescopes, Instruments, and Systems, 1, 027002

\bibitem[{{Torres} {et~al.}(2007){Torres}, {Loinard}, {Mioduszewski}, \&
  {Rodr{\'{\i}}guez}}]{Torres:2007}
{Torres}, R.~M., {Loinard}, L., {Mioduszewski}, A.~J., \& {Rodr{\'{\i}}guez},
  L.~F. 2007, \apj, 671, 1813

\bibitem[{{Torres} {et~al.}(2009){Torres}, {Loinard}, {Mioduszewski}, \&
  {Rodr{\'{\i}}guez}}]{Torres:2009}
---. 2009, \apj, 698, 242

\bibitem[{{van der Marel} {et~al.}(2013){van der Marel}, {van Dishoeck},
  {Bruderer}, {Birnstiel}, {Pinilla}, {Dullemond}, {van Kempen}, {Schmalzl},
  {Brown}, {Herczeg}, {Mathews}, \& {Geers}}]{vandermarel:2013}
{van der Marel}, N., {van Dishoeck}, E.~F., {Bruderer}, S., {et~al.} 2013,
  Science, 340, 1199

\bibitem[{{Voshchinnikov}(1989)}]{Voshchinnikov:1989}
{Voshchinnikov}, N.~V. 1989, Astrophysics, 30, 313

\bibitem[{{Wenzel}(1969)}]{Wenzel:1969}
{Wenzel}, W. 1969, Zentralinstitut fuer Astrophysik Sternwarte Sonneberg
  Mitteilungen ueber Veraenderliche Sterne, 5, 75

\bibitem[{{White} \& {Basri}(2003)}]{White:2003}
{White}, R.~J., \& {Basri}, G. 2003, \apj, 582, 1109

\bibitem[{{Wichmann} {et~al.}(1996){Wichmann}, {Krautter}, {Schmitt},
  {Neuhaeuser}, {Alcala}, {Zinnecker}, {Wagner}, {Mundt}, \&
  {Sterzik}}]{Wichmann:1996}
{Wichmann}, R., {Krautter}, J., {Schmitt}, J.~H.~M.~M., {et~al.} 1996, \aap,
  312, 439

\bibitem[{{Wichmann} {et~al.}(2000){Wichmann}, {Torres}, {Melo}, {Frink},
  {Allain}, {Bouvier}, {Krautter}, {Covino}, \&
  {Neuh{\"a}user}}]{Wichmann:2000}
{Wichmann}, R., {Torres}, G., {Melo}, C.~H.~F., {et~al.} 2000, \aap, 359, 181

\bibitem[{{Yadav} {et~al.}(2015){Yadav}, {Christensen}, {Morin}, {Gastine},
  {Reiners}, {Poppenhaeger}, \& {Wolk}}]{Yadav:2015}
{Yadav}, R.~K., {Christensen}, U.~R., {Morin}, J., {et~al.} 2015, \apjl, 813,
  L31

\bibitem[{{Zacharias} {et~al.}(2004){Zacharias}, {Monet}, {Levine}, {Urban},
  {Gaume}, \& {Wycoff}}]{Zacharias:2004}
{Zacharias}, N., {Monet}, D.~G., {Levine}, S.~E., {et~al.} 2004, in Bulletin of
  the American Astronomical Society, Vol.~36, American Astronomical Society
  Meeting Abstracts, 1418

\bibitem[{{Zhou} {et~al.}(2016){Zhou}, {Rodriguez}, {Collins}, {Beatty},
  {Oberst}, {Heintz}, {Stassun}, {Latham}, {Kuhn}, {Bieryla}, {Lund},
  {Labadie-Bartz}, {Siverd}, {Stevens}, {Gaudi}, {Pepper}, {Buchhave},
  {Eastman}, {Col{\'o}n}, {Cargile}, {James}, {Gregorio}, {Reed}, {Jensen},
  {Cohen}, {McLeod}, {Tan}, {Zambelli}, {Bayliss}, {Bento}, {Esquerdo},
  {Berlind}, {Calkins}, {Blancato}, {Manner}, {Samulski}, {Stockdale},
  {Nelson}, {Stephens}, {Curtis}, {Kielkopf}, {Fulton}, {DePoy}, {Marshall},
  {Pogge}, {Gould}, {Trueblood}, \& {Trueblood}}]{Zhou:2016}
{Zhou}, G., {Rodriguez}, J.~E., {Collins}, K.~A., {et~al.} 2016, \aj, 152, 136

\bibitem[{{Zimmermann} {et~al.}(1993){Zimmermann}, {Belloni}, {Izzo},
  {Kahabka}, \& {Schwentker}}]{Zimmermann:1993}
{Zimmermann}, H.~U., {Belloni}, T., {Izzo}, C., {Kahabka}, P., \& {Schwentker},
  O. 1993, {EXSAS user's guide. Extended Scientific Analysis System to evaluate
  data from the astronomical X-ray satellite ROSAT.}

\end{thebibliography}

\end{document}